\documentclass[a4paper,11pt]{article}
\pdfoutput=1 
\usepackage{afterpage}
\usepackage{jcappub} 
\usepackage{fontenc} 
\usepackage{tabularx}
\usepackage{hhline}
\usepackage{graphicx,epstopdf}
\usepackage{float}
\usepackage{afterpage}
\usepackage{verbatim}
\usepackage{booktabs}
\usepackage[section]{placeins}
\usepackage{amssymb}
\usepackage{multirow,bigdelim}
\DeclareGraphicsExtensions{.pdf,.png,.jpg}

\newcommand{\ra}[1]{\renewcommand{\arraystretch}{#1}}

\title{Investigating the Uniformity of the Excess Gamma rays towards the Galactic Center Region}

\author[a]{Shunsaku Horiuchi,}
\author[b]{Manoj Kaplinghat,}
\author[b]{and Anna Kwa}

\affiliation[a]{Center for Neutrino Physics, Department of Physics, Virginia Tech, Blacksburg, Virginia 24061, USA}
\affiliation[b]{Center for Cosmology, Department of Physics and Astronomy, University of California, Irvine, Irvine, California 92697 USA}

\emailAdd{horiuchi@vt.edu}
\emailAdd{mkapling@uci.edu}
\emailAdd{akwa@uci.edu}

\abstract{
We perform a composite likelihood analysis of subdivided regions within the central $26^\circ\times20^\circ$ of the Milky Way, with the aim of characterizing the spectrum of the gamma-ray galactic center excess in regions of varying galactocentric distance. Outside of the innermost few degrees, we find that the radial profile of the excess is background-model dependent and poorly constrained. The spectrum of the excess emission is observed to extend upwards of 10 GeV outside $\sim5^\circ$ in radius, but cuts off steeply between 10--20 GeV only in the innermost few degrees. If interpreted as a real feature of the excess, this radial variation in the spectrum has important implications for both astrophysical and dark matter interpretations of the galactic center excess. Single-component dark matter annihilation models face challenges in reproducing this variation; on the other hand, a population of unresolved millisecond pulsars contributing both prompt and secondary inverse Compton emission may be able to explain the spectrum as well as its spatial dependency. We show that the expected differences in the photon-count distributions of a smooth dark matter annihilation signal and an unresolved point source population are an order of magnitude smaller than the fluctuations in residuals after fitting the data, which implies that mismodeling is an important systematic effect in point source analyses aimed at resolving the gamma-ray excess.
}

\begin{document}
\maketitle
\flushbottom

\section{Introduction}
\label{sec:intro}
Fermi Large Area Telescope (LAT) observations towards the Milky Way center have revealed a spatially extended source of gamma rays in excess of the modeled astrophysical backgrounds \cite{Goodenough:2009gk,Vitale:2009hr, Hooper:2010mq,Hooper:2011ti, Abazajian:2012pn,Gordon:2013vta,Macias:2013vya,Hooper:2013rwa, Abazajian:2014fta,Abazajian:2014hsa,Zhou:2014lva, Daylan:2014rsa,Calore:2014xka,TheFermi-LAT:2015kwa}. This `galactic center excess' (GCE) has so far been found to be robust against variations in background modeling \cite{Cumberbatch:2010ii,Macias:2013vya,Zhou:2014lva, Abazajian:2014fta,Calore:2014xka,TheFermi-LAT:2015kwa}. Possible explanations for this excess include weakly-interacting massive particle (WIMP) dark matter annihilations, unresolved milllisecond pulsars (MSPs), and cosmic-ray outbursts from the galactic center. 

The interpretation of the GCE as emission from dark matter annihilations has raised considerable interest due to the findings of Refs. \cite{Goodenough:2009gk,Abazajian:2012pn,Gordon:2013vta,Macias:2013vya,Abazajian:2014fta,Daylan:2014rsa,Abazajian:2014hsa,Calore:2014xka,TheFermi-LAT:2015kwa} that point out (1) the spatial morphology of the GCE is consistent with that WIMP annihilations in a Navarro-Frenk-White (NFW) dark matter halo, (2) the spectrum of the GCE is consistent with predictions for WIMP annihilations into Standard Model particles, and (3) the annihilation cross sections required to fit the modeled spectra to the data are of the same order as the weak-scale annihilation cross section that results in the observed relic abundance of dark matter. However, the excess emission may also be attributed to astrophysical sources. 
A large unresolved population of millisecond pulsars remains a viable astrophysical explanation for the excess emission~\cite{Abazajian:2012pn,Mirabal:2013rba,Gordon:2013vta,Petrovic:2014xra,Yuan:2014rca,Yuan:2014yda,Brandt:2015ula,O'Leary:2015gfa,O'Leary:2016osi}: the typical MSP spectrum, as observed in globular clusters, is consistent with the observed GCE spectrum \cite{Abazajian:2010zy}, and the spatial distribution of low-mass X-ray binaries (which are thought to be an earlier evolutionary phase of MSPs) is consistent with an NFW-like power law, at least in M31~\cite{Yuan:2014rca}.\footnote{The population in M31 is used instead of the Milky Way as the current INTEGRAL catalog of low-mass X-ray binaries in the Milky Way bulge has substantial completeness concerns~\cite{Bodaghee:2007db}.} 
Additionally, the central regions of the Milky Way have experienced violent eruptions in the past, as evidenced by the lobed structures of the Fermi bubbles emanating from the galactic center \cite{Su:2010qj}; this history of burst activity has motivated authors to consider cosmic-ray injection events as another possible astrophysical explanation for the GCE \cite{Carlson:2014cwa,Petrovic:2014uda,Gaggero:2015nsa, Cholis:2015dea}.

Any spectral or spatial variation (or lack thereof) in the signal would be of critical importance in discerning amongst the possible origins of the GCE. For example, a prompt dark matter annihilation signal---where the subsequent decay and hadronization of the Standard Model products occurs quickly and the ensuing gamma rays are emitted at the site of annihilation---should have an intensity directly proportional to the square of the dark matter density profile, and the spectrum should be independent of sky position. On the other hand, if the GCE was at least partially produced through inverse Compton (IC) scattering from a population of high-energy leptons---which is possible in the cases of MSPs \cite{Petrovic:2014xra, Yuan:2014yda}, dark matter annihilations to leptons \cite{Lacroix:2014eea,Calore:2014nla,Kaplinghat:2015gha,Lacroix:2015wfx}, or a leptonic cosmic-ray outburst---its spectrum and intensity would be dependent on cosmic-ray diffusion processes in the central Milky Way as well as the interstellar radiation field (ISRF). We might thus expect to observe some variation in the spectral shape and normalization as a function of sky position if the GCE source was (at least partially) leptonic.\footnote{Hadronic cosmic-ray outbursts may also impinge upon gas and produce gamma rays through subsequent $\pi^0$ decays and bremsstrahlung processes; however, the gamma rays produced in this scenario will trace the gas distribution and thus have a disk-like, not spherical, morphology.} 

The GCE has been observed within the innermost few degrees of the Milky Way \cite{Goodenough:2009gk,Abazajian:2012pn,Gordon:2013vta,Macias:2013vya,Abazajian:2014fta,Daylan:2014rsa,Abazajian:2014hsa} (hereafter referred to as the `galactic center') as well as the region immediately exterior to the galactic center (hereafter referred to as the `inner galaxy') \cite{Daylan:2014rsa,Calore:2014xka}. In this paper, we compare the best-fit spectra and morphologies across multiple regions, including the galactic center, using consistent diffuse background models and fitting procedures between analyses of each region. The key idea is to use the morphologies of the diffuse backgrounds to constrain the spectrum and thereby investigate the spatial uniformity and photon-count distribution of the excess. We discuss our results in terms of their implications for both dark matter and astrophysical interpretations of the GCE. In particular, we investigate (1) the presence of a power law-like feature in the GCE spectrum, with emission extending upwards of $\sim20$ GeV \cite{Calore:2014xka, TheFermi-LAT:2015kwa} and (2) the consistency of the GCE with a population of unresolved MSPs \cite{Bartels:2015aea, Lee:2015fea}.

Analyses of the excess in the inner galaxy report a power law-like high-energy tail in the GCE  spectrum beyond 10 GeV, possibly extending upwards of $\sim20-100$ GeV \cite{Calore:2014nla, Calore:2014xka, TheFermi-LAT:2015kwa}. If the GCE is a bona fide signal from dark matter annihilations, multiple particle properties (mass, annihilation primaries, branching ratios) are encoded within the shape of its gamma-ray spectrum. Inclusion or exclusion of the high-energy tail in the GCE spectrum can greatly affect the best fit dark matter mass and annihilation channel(s). If the GCE has an astrophysical origin, the presence of high-energy emission could inform us about the processes that gave rise to it and perhaps rule out certain scenarios. In Sections \ref{subsec:spatialGCE} and \ref{subsec:spectrumtail} we investigate whether the high-energy tail of the GCE spectrum originates from the same source that produces its spectrum below $\sim$10 GeV. 

Recent results support an unresolved point source origin for the GCE and indicate that such sources may be able to account for the entirety of the excess in the inner galaxy \cite{Lee:2015fea, Bartels:2015aea}. If so, this would strongly imply that most, if not all, of the GCE signal is produced by millisecond pulsars, not dark matter annihilation. This would be evidence of an as-yet-undiscovered pulsar population at the galactic center with exciting implications for astronomy across the electromagnetic spectrum. The MSP interpretation of the excess would also set strong upper limits on the WIMP annihilation cross section. In Section \ref{subsec:ptsrcs} we attempt to determine whether our findings are suggestive of either an unresolved point source distribution or annihilation in a smooth NFW halo.

\section{Methods}
\label{sec:methods}
\begin{figure}[H]
 \centering
 \includegraphics[width=6.truein]{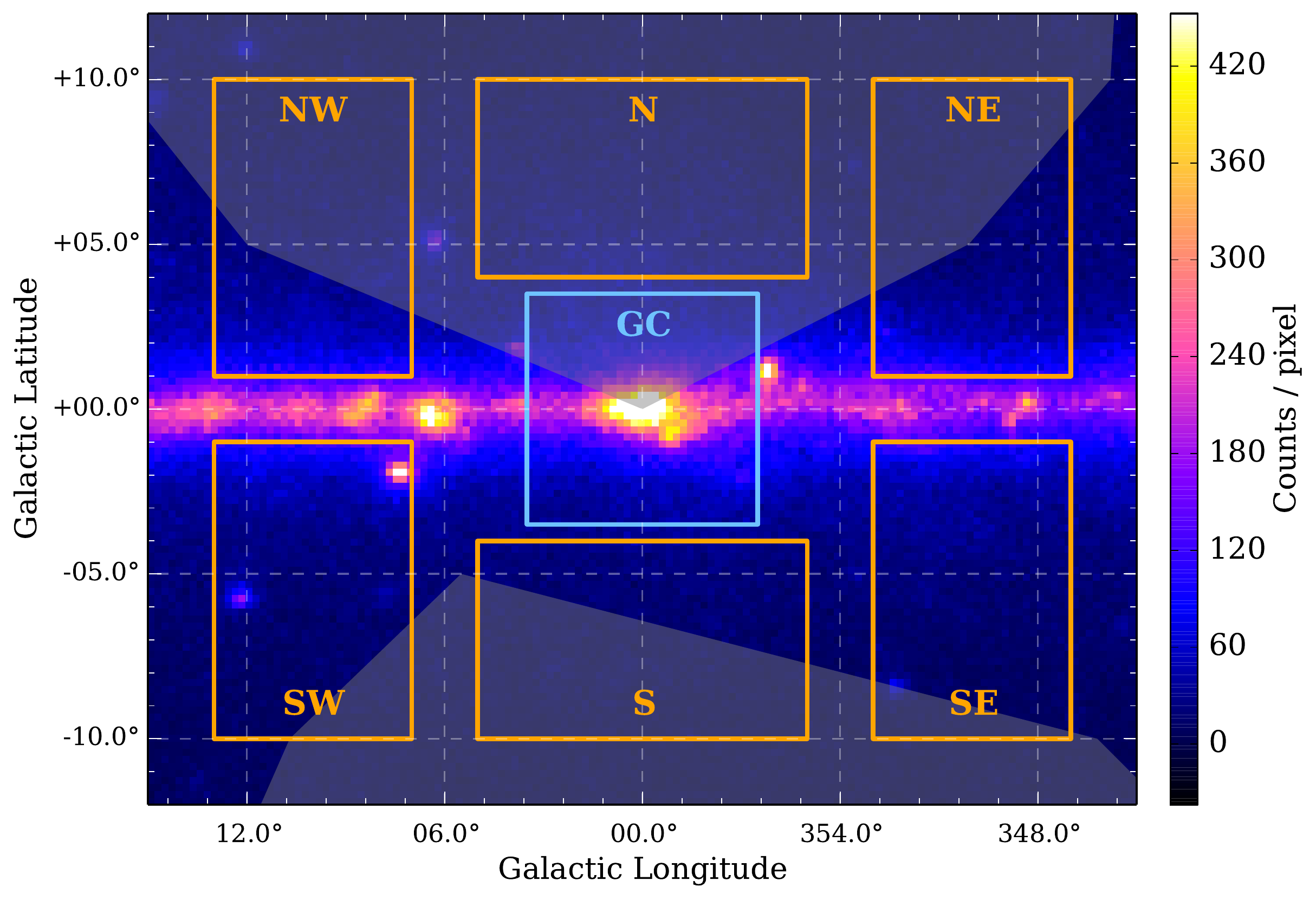}
 \caption{Total observed counts map with labelled regions of interest. Each ROI will be referred to hereafter using its label from this figure. We will collectively refer to the entirety of the ROIs \textit{excluding} the innermost $7^\circ\times7^\circ$ as the `inner galaxy' (orange). The innermost $7^\circ\times7^\circ$ ROI is referred to as the `galactic center' (blue). The overlap of the Fermi bubble template used in this analysis (c.f. Sec.~\ref{subsec:fitcomponents}) with the ROIs is shown as the gray overlay. Not shown here are the farther latitude N2/S2/bubble N/bubble S ROIs which were used to constrain the Fermi bubble spectrum and estimate the extent of the GCE signal.}
  \label{fig:ROIs}
\end{figure}

We use approximately 73 months of Pass 7 data from the \textit{Fermi-LAT} taken between August 2008 and September 2014.\footnote{This study uses the Pass 7 data as it was commenced before the public release of the Pass 8 dataset. Ref.~\cite{Linden:2016rcf} show that using Pass 7 versus Pass 8 data has a negligible effect on the GCE spectrum, including the higher energies.} We use \texttt{CLEAN}-class photon events and the Pass 7 reprocessed instrument response functions. The photon events range from 700 MeV--200 GeV and are binned in 8 logarithmically spaced bins from 700 MeV--10 GeV and 3 high-energy bins from 10--200 GeV. We use larger bin sizes above 10 GeV to compensate for the lower photon counts at high-energies. We apply a maximum zenith angle cut of $90^\circ$ to avoid contamination from Earth limb emission. We define our `inner galaxy' regions of interest (ROIs) and `galactic center' ROIs in Tab. \ref{tab:ROIs} as well as Fig. \ref{fig:ROIs}.

Many studies of the galactic center emission, including this work, rely on spatial template-based analyses in which the GCE spectrum is fit alongside the spectra of the background diffuse emission and point sources. The strengths of such analyses lie in their ability to effectively subtract out the bulk of the astrophysical backgrounds from the data. However, this method inherently introduces systematic effects by assuming a given spatial profile for each diffuse background source. Potentially large errors in the best-fit GCE spectrum may arise if the spatial templates assumed for the backgrounds differ considerably from the true background emission. For this reason we test three different diffuse background models (described below in Sec.~\ref{subsec:fitcomponents}) to estimate the systematic error in the GCE spectrum due to the uncertainty in our assumptions about the astrophysical background. Our results are valid and robust under the three models tested below, but note that the range of backgrounds tested here is more limited than used in previous works~\cite{Zhou:2014lva,Calore:2014xka}.

\subsection{Fit components}
\label{subsec:fitcomponents}
The gamma-ray observations are modeled as a combination of the following source templates: \\

\textbf{Diffuse gamma-ray background: }The primary diffuse astrophysical gamma-ray background is produced by the following processes:
\begin{itemize}
 \item Neutral pion ($\pi^0$) decay: Neutral pions are produced when hadronic cosmic rays impinge upon clouds of gas in the interstellar medium (ISM). 	The $\pi^0$'s subsequently decay into pairs of high-energy photons.
 \item Bremsstrahlung radiation: High-energy electron cosmic rays interact with gas in the ISM.
 \item Inverse Compton radiation: High-energy electron cosmic rays upscatter lower energy background starlight photons in the interstellar radiation field.
\end{itemize}

We model the gamma-ray emission from the above processes using the WebRun interface of the GALPROP (version 54) cosmic-ray propagation code \cite{galprop, Porter:2008ve, Vladimirov:2010aq}, which computes the diffusion and energy losses for a chosen set of propagation parameters and outputs the resultant gamma-ray skymap templates and spectra. 
For a given diffusion model, we generate the emission templates from $\pi^0$ decay and bremsstrahlung separately and then combine them into a single $\pi^0+$bremsstrahlung diffuse template.
The spatial distribution of background emission from $\pi^0$ decay and bremsstrahlung radiation is very similar because both processes require the same gas cloud target. If individual templates are included for the $\pi^0$ and bremsstrahlung emission, the large degeneracies between the two spatial morphologies would make it difficult for the likelihood maximization to correctly fit the spectrum of each component. We therefore fit a single, combined $\pi^0+$bremsstrahlung diffuse template in each energy bin to avoid the inclusion of two templates with largely degenerate morphologies. 
The $\pi^0$+bremsstrahlung and IC components are fit independently of each other. 

To test the robustness of our results, we repeat the analysis with three different diffuse gamma-ray background models generated using GALPROP. We use models selected from the suite of diffuse backgrounds tested by Ref.~\cite{Calore:2014xka} in their systematic analysis of the GCE signal. For consistency and ease of comparison between works, we refer to the background models using the same labelling (A/E/F) as in Ref.~\cite{Calore:2014xka}. The variations in the input parameters for our diffuse backgrounds are listed in Tab.~\ref{tab:galpropmodels}. GALPROP  model A is chosen for testing as it is `tuned' such that the recovered best-fit template normalizations after fitting to the data agree well with the GALPROP prediction. Model F is chosen as it was found to provide the highest likelihood fit in the inner galaxy between $2^\circ<|b|<20^\circ$. We chose to test GALPROP model E as an extreme case: the low diffusion coefficient $D_0$ in this model leads to a large bump in the IC spectrum below 10 GeV as well as different spatial morphologies compared to models A and F. The effects of fitting with this extreme background are further discussed in Sec.~\ref{subsec:bgmodelresults}. For a detailed description of the effects of varying diffuse model parameters on the characterization of the GCE, see Refs.~\cite{Calore:2014xka, TheFermi-LAT:2015kwa}. 

\setlength{\tabcolsep}{4.5pt}
\begin{table*}[t]
\centering
\begin{tabular}{cccccccccc}
\hline\\ 

 Model &  $z_D  $ & $D_0 $  & $dv/dz$ &  CR Source & $\alpha_e / \alpha_p$ & $N_e/N_p$  & B-field &  ISRF & $T_S$   \\ \hline
\\
A  &  4 &  5.0 & 50 & SNR     & 2.43/2.47 & 2.00/5.8 & 090050020  & 1.36/1.36/1.0 & 150     \\
E  &  4 &  2.0 & 0  & SNR     & 2.43/2.39 & 0.40/4.9 & 050100020  & 1.0/1.0/1.0   & 150     \\
F  &  6 &  8.3 & 0  & PLS$_L$ & 2.42/2.39 & 0.49/4.8 & 050100020  & 1.0/1.0/1.0   & $10^5$  \\

\hline
\end{tabular}
\caption{Input parameters for our set of three GALPROP diffuse background models. We use the same scale radius $r_D$=20 kpc and Alfv\'{e}n speed $v_A$=32.7 km s$^{-1}$ for all models. The scale height $z_D$ is given in units of kpc. The diffusion coefficient $D_0$ is given in units $\times10^{28} $cm$^3$ s$^{-1}$. The convection velocity gradient $dv/dz$ is given in km s$^{-1}$ kpc$^{-1}$. The cosmic-ray source distribution is taken from either the measured supernova remnant (SNR) distribution \cite{Case:1998qg} or the Lorimer pulsar distribution \cite{Lorimer:2006qs}. (Both of these cosmic-ray distributions approach zero at the galactic center and are in all likelihood severely underestimating the cosmic-ray source density in the innermost kpc. We discuss the implications of this deficiency in Sec.~\ref{subsec:bgmodelresults}.) The power law index of the electron (proton) injection spectrum above rigidity 2.18 (11.3) GV is given by $\alpha_e (\alpha_p)$. The electron (proton) cosmic-ray injection spectrum is normalized to $N_e (N_p)$ in units of $\times10^{-9}$ cm$^{-2}$ sr$^{-1}$ s$^{-1}$ MeV$^{-1}$ at 34.5 (100) GeV. The first set of three digits in the magnetic field model are $B_0\times10 \mu$G, the second set of three digits are $r_c\times10$ kpc, and the last set of three digits are $z_c\times10$ kpc. ISRF normalization factors are given for the optical, IR, and CMB components respectively. The gas spin temperature $T_S$ is in units of K. A fuller description of the parameters may be found in Refs.~\cite{FermiLAT:2012aa, Carlson:2016iis}.  }
\label{tab:galpropmodels} 
\end{table*}

\textbf{GCE template: } The GCE is well-fit by annihilation signals based on NFW profiles \cite{Navarro:1995iw}, which approximate cold dark matter halo densities in N-body simulations. We therefore base our set of GCE templates upon the signal morphology that is predicted for annihilations in an NFW halo, which is proportional to the density squared, integrated along the line of sight. It should be noted that the spatial profile we assume in our template model for the GCE is not unique to dark matter annihilations, and may also be consistent with a central MSP population \cite{Abazajian:2012pn}. The dark matter density profiles in N-body simulations have been found to follow the functional form 
\begin{equation}
 \rho(r)=\frac{\rho_s}{(r/r_s)^\gamma [1+r/r_s]^{(3-\gamma)}} ~~.
\end{equation}
The density profile is normalized to the local dark matter density at the solar position, $\rho_\odot=0.3$ GeV cm$^{-3}$ \cite{Zhang:2012rsb}, with scale radius $r_s=23$ kpc. The log slope of the NFW density profile asymptotes to the inner slope $\gamma$ as $r$ approaches the halo center.
In regions lying outside the central few degrees, such as our inner galaxy ROIs, the density slope begins to deviate from the asymptotic inner value $\gamma$. Thus, there is some degeneracy between the NFW inner profile slope $\gamma$ and the scale radius $r_s$ as the region of interest moves away from the galactic center. In outlying regions (but not the galactic center), a GCE template with a shallower inner slope $\gamma$ and smaller scale radius $r_s$ may be similar in morphology to a template with a steeper inner slope and larger scale radius. Since our aim is to describe the morphology preferred by the inner galaxy excess (rather than infer the parameters of the NFW-like profile), we fix the scale radius to $r_s=23$ kpc.

For a source originating from Majorana dark matter annihilations with velocity-averaged cross section $\langle\sigma v\rangle$, the differential flux  received along a line of sight towards galactic coordinates $(l,b)$ is given by
\begin{equation}
 \frac{d\Phi(l,b)}{dE} = \frac{1}{4 \pi m_\chi^2} \frac{\langle\sigma v\rangle}{2} J(l,b) \frac{dN_\gamma}{dE} ~~,
\end{equation}
where $m_\chi$ is the dark matter particle mass, and $dN_\gamma/dE$ is the gamma-spectrum per annihilation. The quantity $J(l,b)$, commonly referred to as the `J-factor', depends on the astronomical dark matter distribution and is equal to the mass density squared, integrated over the line of sight $x$ through $(l,b)$:
\begin{equation}
 J=  \int_{l.o.s}^{} \rho^2(r_{GC}(x,l,b))~dx 
\end{equation}
where $r_{GC}=[R_{\odot}^2-2xR_\odot \text{cos}(l)\text{cos}(b)+x^2]^{1/2}$ is the distance from the galactic center. For our ROIs close to the Milky Way center, $r \ll r_s$ so that $\rho(r)\propto r^{-\gamma}$. We use the value $R_\odot=8.25$ kpc for the solar distance to the galactic center.

As previously discussed, one of our goals is to test whether the GCE in the galactic center and inner galaxy regions can be described with a single NFW annihilation profile. We use NFW annihilation templates with inner slopes $\gamma=\{0.9, 1.0, 1.1, 1.2, 1.3\}$ when fitting the GCE in the galactic center and inner galaxy to test whether any of these profiles yields consistent fluxes between various ROIs. We also test a template with $\gamma=0.8$ for the GALPROP model F background in the inner galaxy, which is the one case we find where the likelihood favors shallower NFW profiles.

\textbf{Fermi bubbles: } The Fermi bubbles are a diffuse, lobed source extending up to $\sim50^\circ$ North and South in latitude from the galactic center \cite{Su:2010qj}. The bubbles are found by Ref.~\cite{Fermi-LAT:2014sfa} to have a hard spectrum of $dN/dE\propto E^{-1.9\pm0.2}$ with a high-energy cutoff around 100 GeV. We employ a flat emission template with edges defined as in Ref.~\cite{Su:2010qj} because our GALPROP-generated diffuse backgrounds do not model this extended source. The use of a uniform spatial template is motivated by Refs.~\cite{Su:2010qj, Fermi-LAT:2014sfa}'s finding that the bubbles' intensity is approximately flat in projection. As the bulk of the Fermi bubbles' emission lies at farther latitudes outside our ROIs, we use additional regions defined by $330^\circ\leq l \leq 20^\circ$, $20^\circ \leq |b| \leq 35^\circ $ in the northern and southern galactic hemispheres to externally constrain their spectrum within the ROI. These regions were chosen to lie outside of $b=20^\circ$ to avoid overlap with other subregions. Although the northern lobe of the bubble template extends into our galactic center ROI, we do not include this template as part of the galactic center fit as (1) its spatial profile becomes uncertain at low latitudes and (2) its flux per steradian is subdominant to other extended components in the galactic center.

\textbf{20 cm gas template: }We include a gas template for the galactic ridge structure as previously described in Refs.~\cite{Aharonian:2006au, YusefZadeh:2012nh, Macias:2013vya, Abazajian:2014fta, Abazajian:2014hsa}. This emission, which is correlated with 20 cm radio emission as well as $\sim$TeV gamma-ray emission in the central region, has been interpreted by Refs.~\cite{YusefZadeh:2012nh, Macias:2013vya, Abazajian:2014fta, Abazajian:2014hsa} as bremsstrahlung emission from a population of high-energy electrons in the galactic center.

\textbf{WISE 3.4 $\mu$m template: }We include a template tracing the infrared starlight emission in the galactic center. This component was interpreted in Ref.~\cite{Abazajian:2014hsa} to be IC emission (in excess of the GALPROP predicted IC flux) from a population of high-energy leptons. We do not find any flux associated with this template in our analysis, which is consistent with our previous finding in Ref.~\cite{Abazajian:2014hsa} which found that the spectrum of the component had a steep cutoff around $\sim$400--500 MeV, which is below the minimum energy of our analysis. This is also found in the work of Ref.~\cite{Lacroix:2015wfx} using self-consistent GALPROP modeling of the additional IC emission. Thus, we do not show any spectrum for this component in Figs.~\ref{fig:bgspectra} and \ref{fig:bgspectra_fulldiffusetied}.

\textbf{Point sources: }We include point sources from the Fermi 3FGL point source catalog \cite{Acero:2015gva} that lie within or near the regions of interest (ROIs). Sources listed in the catalog with significance $\sigma > 5.0$ are free to have their spectra varied.

\textbf{Isotropic gamma-ray background: }We do not assume a fixed spectrum for the isotropic extragalactic background, but fit the normalization of a uniform isotropic background template independently in each energy bin. We note that Refs.~\cite{Abazajian:2014fta, Daylan:2014rsa, Abazajian:2014hsa} find evidence for an isotropic or close to isotropic component in the innermost $7^\circ\times7^\circ$ that is somewhat brighter than the Fermi collaboration's standard extragalactic isotropic background template \cite{Ackermann:2014usa}. Refs.~\cite{Abazajian:2014fta, Abazajian:2014hsa} fit this component with either an additional isotropic template (Ref.~\cite{Abazajian:2014fta}) or a nearly isotropic `new diffuse' (ND) template (Ref.~\cite{Abazajian:2014hsa}). Hence, we test the case where the isotropic component in the galactic center region is allowed to vary separately from the isotropic component in the inner galaxy. 

\vspace{5mm}

\begin{table*}[h]
\centering
\begin{tabular}{lccc}

\toprule \\
Region of interest &  Range in $l$   &  Range in $b$   & Angular area (sr)  \\ \hline \\
\textbf{Galactic center} & & & \\
 ~~GC           & $-3.5^\circ \leq l \leq 3.5^\circ$  & $-3.5^\circ \leq b \leq 3.5^\circ $ &   $1.49\times10^{-2}$  \\ \hline \\ 
\textbf{Inner galaxy} & & & \\
 ~~N            & $-5^\circ \leq l \leq 5^\circ $ &  $4^\circ \leq b \leq 10^\circ $     &   $1.83\times10^{-2}$     \\
 ~~S            & $-5^\circ \leq l \leq 5^\circ $ &  $-10^\circ \leq b \leq -4^\circ $     &   $1.83\times10^{-2}$  \\
 ~~NE            & $347^\circ \leq l \leq 353^\circ $  & $ 1^\circ \leq b \leq 10^\circ $  &   $1.64\times10^{-2}$    \\
 ~~NW           & $7^\circ \leq l \leq 13^\circ $  & $ 1^\circ \leq b \leq 10^\circ $ &  $1.64\times10^{-2}$  \\ 
 ~~SE           & $347^\circ \leq l \leq 353^\circ $  & $ -10^\circ \leq b \leq -1^\circ $  &   $1.64\times10^{-2}$  \\
 ~~SW          & $7^\circ \leq l \leq 13^\circ $  & $ -10^\circ \leq b \leq -1^\circ $  &  $1.64\times10^{-2}$    \\  \hline \\ 
 ~~N2		& $-5^\circ \leq l \leq 5^\circ $  & $ 11^\circ \leq b \leq 19^\circ $  & $2.44\times10^{-2}$ \\
 ~~S2		& $-5^\circ \leq l \leq 5^\circ $  & $ -19^\circ \leq b \leq -11^\circ $  & $2.44\times10^{-2}$ \\
 ~~bubble N    & $330^\circ \leq l \leq 20^\circ $  & $ 25^\circ \leq b \leq 35^\circ $  &  $2.28\times10^{-1}$   \\ 
 ~~bubble S    & $330^\circ \leq l \leq 20^\circ $  & $ -35^\circ \leq b \leq -25^\circ $  &  $2.28\times10^{-1}$   \\  
    
 \bottomrule
\end{tabular}
\caption{Our regions of interest, as defined by range in galactic longitude $l$ and latitude $b$. Angular areas in steradians are also given, although all our results for best-fit GCE flux in each ROI are normalized to display the total expected flux (GeV s$^{-1}$ cm$^{-2}$) from the 35$^\circ\times35^\circ$ GCE template, based on the observed flux for each individual ROI. Note that the farthest latitude regions `bubble N/S' and `N2/S2' are not included in the GCE analysis and thus not shown in Fig.~\ref{fig:ROIs}; they are included solely for the purpose of constraining the Fermi bubbles' spectrum and testing the extent of the GCE signal. }
\label{tab:ROIs} 
\end{table*}

\subsection{Fit procedure}
\label{subsec:compositeanalysis}
We use the \textit{Composite2} tool within the Fermi Science Tools Python interface \cite{fermitools} to perform a composite likelihood analysis of multiple ROIs simultaneously for each energy bin. This allows for any number of chosen model parameters---e.g. flux normalization of the diffuse background components in the chosen energy bin---to be tied across multiple ROIs, while still allowing for the possibility that other extended sources---e.g. the GCE template---might be fit with different normalizations between ROIs. We constrain the normalization of the extended astrophysical sources (GALPROP $\pi^0$+bremsstrahlung diffuse, GALPROP IC diffuse, Fermi bubbles, and isotropic background templates) to be the same throughout all ROIs in Tab.~\ref{tab:ROIs}. The origin of the GCE is yet unknown and we do not presume that it must be fit with a single spectrum and template normalization across all regions. Thus, we allow the GCE template to be fit with different normalizations in the individual inner galaxy and galactic center ROIs shown in Fig.~\ref{fig:ROIs}.  

We perform purely spatial fits to the data within each independent energy bin, i.e., we do not require the modeled sources to follow any fixed spectral shape or parameterized functional form across multiple energy bins. This is also true for the galactic diffuse templates and isotropic background, which are typically constrained to have a fixed spectral shape. We note that the GALPROP code does give a prediction of the spectrum for each diffuse background component; however, we do not constrain the normalization of the diffuse templates to follow the GALPROP-predicted spectral shapes when fitting.

Our choice of methodology does entail some caveats. As previously mentioned, the $\pi^{0}$+bremsstrahlung and IC diffuse backgrounds are not fixed to the broadband spectral shapes predicted by GALPROP for each of the models. Therefore, the best-fit spectrum for either of these diffuse components may be unphysical in the sense that it does not necessarily correspond to the GALPROP parameters that produce the spatial profile it is associated with. In principle, fitting the background in independent energy bins allows more freedom for the diffuse backgrounds to absorb the GCE component. However, we find that for extreme background model parameters---such as in model E, where the low diffusion coefficient leads to a large modification in the IC component---the modeled background is a poor fit to the actual gamma-ray diffuse background, which causes the fitting procedure to lower the normalization of the background model in favor of increasing the normalization of the GCE or other extended templates (see Sec.~\ref{subsec:bgmodelresults}). We also note that this analysis does not include a template for the large scale feature Loop I in the northern galactic sky. If this omission affected the derived GCE spectrum in the inner galaxy, we would expect to observe lower intensity in the best-fit GCE spectrum in our northern ROIs relative to the south; however, we show in Fig.~\ref{fig:IG_ROIs_3x3} that this is not the case.

\section{Results}

In Sec.~\ref{subsec:bgmodelresults} we describe the systematic variations in the best-fit GCE spectra associated with the diffuse background model components, and in particular, the GALPROP-generated IC templates. Sec.~\ref{subsec:GCvsIGspectrum} compares and contrasts the best-fit GCE spectra in the galactic center and combined inner galaxy regions. In Sec.~\ref{subsec:gammas} we discuss how the choice of GALPROP background model affects the best-fit NFW slope and the GCE residual radial profile. Finally, we examine the spatial uniformity of the GCE across the separate ROIs in Sec.~\ref{subsec:spatialGCE}.

\label{sec:results}
\subsection{Systematics associated with background model components}
\label{subsec:bgmodelresults}

\begin{figure}[h]
 \centering
\includegraphics[width=6.2truein]{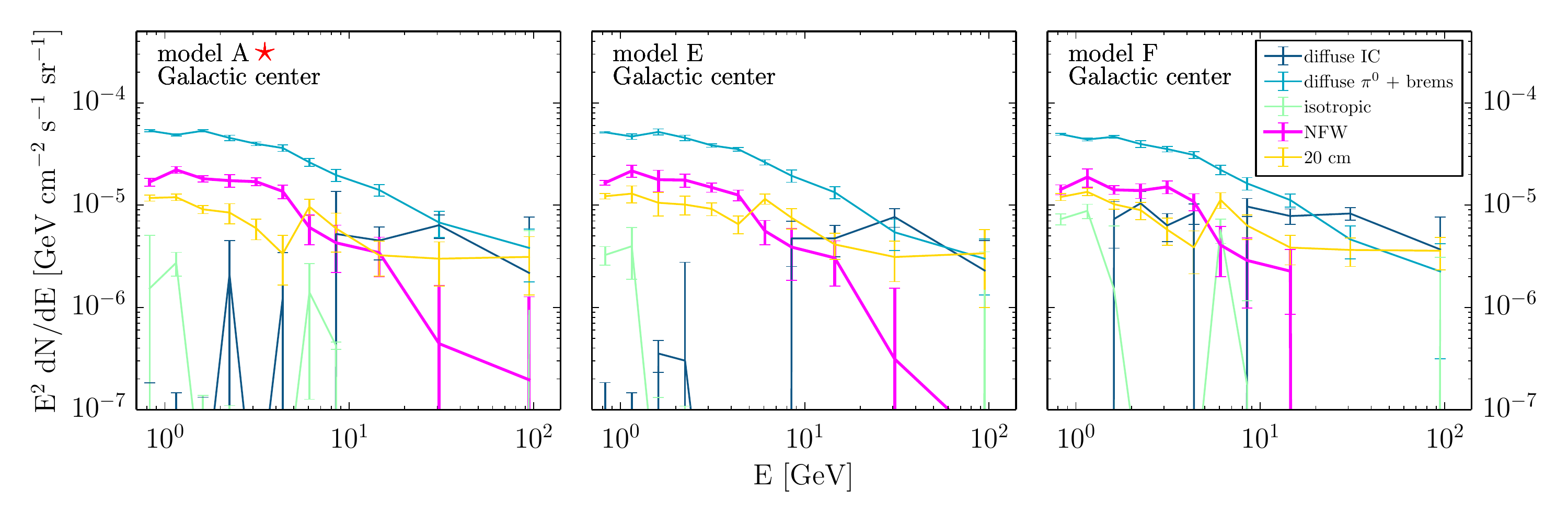}\\  \vspace{-3.5mm}
\includegraphics[width=6.2truein]{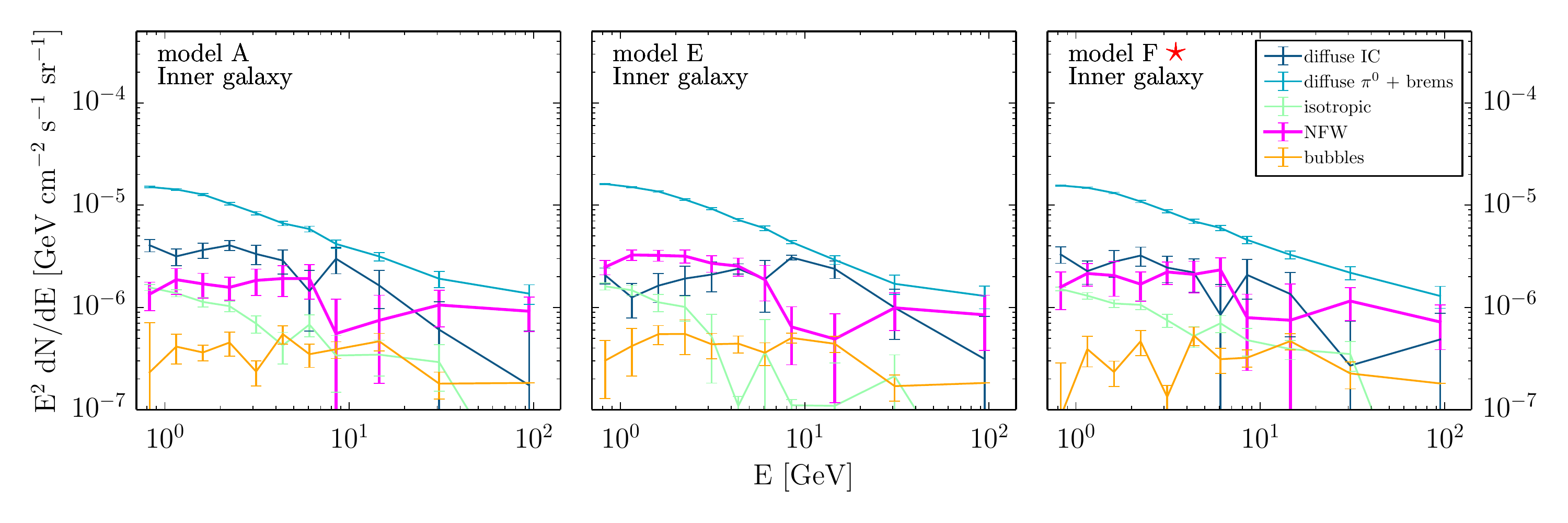}\\  \vspace{-5mm}
\caption{Spectra for the GCE and spatially extended background model components in the galactic center (top row) and combined inner galaxy ROIs (bottom row). Fits were performed with the GALPROP IC and $\pi^0$+bremsstrahlung templates free to vary independently of each other in each energy bin. Red $\bigstar$ symbols denote the best fitting background model in the respective regions.}
\label{fig:bgspectra}
\end{figure}

Fig.~\ref{fig:bgspectra} shows the spectra of the NFW template and diffuse background model components for the three different GALPROP model backgrounds. GCE spectra are shown for the cases where the excess was fit using an NFW template with slope  $\gamma=1.1$, which was found to be the favored value of $\gamma$ in all but one of the fits (see Sec.~\ref{subsec:gammas}). From the upper row of Fig.~\ref{fig:bgspectra} we see that when the spectral shapes of the isotropic and IC diffuse backgrounds are not fixed (but are instead allowed to vary in normalization in independently-fit energy bins), these components are severely under-fit in the galactic center ROI. This indicates that the generic diffuse background models calculated by the GALPROP code are not able to adequately model the spatial distribution of diffuse emission in the innermost $\sim$kpc of the galaxy.

The default spatial distributions for cosmic-ray injection used to model the IC emission in GALPROP are peaked between r$\sim2-5$ kpc (depending on the model used). As pointed out by Refs.~\cite{Gaggero:2015nsa,Carlson:2015ona,Carlson:2016iis}, insertion of a strong source of cosmic rays at the galactic center affects the diffuse background modeling and could thus also affect the derived characteristics of the gamma-ray excess. Using a specialized, local model for the IC emission close to the galactic center, the Fermi collaboration \cite{TheFermi-LAT:2015kwa} finds that this component is strongly enhanced relative to previous diffuse background models; this suggests that the spatial models of cosmic-ray lepton injection in GALPROP are deficient within the innermost kpc and do not produce an accurate representation of the IC emission there. 

\vspace{5mm}

\begin{figure}[h]
 \includegraphics[width=6.2truein]{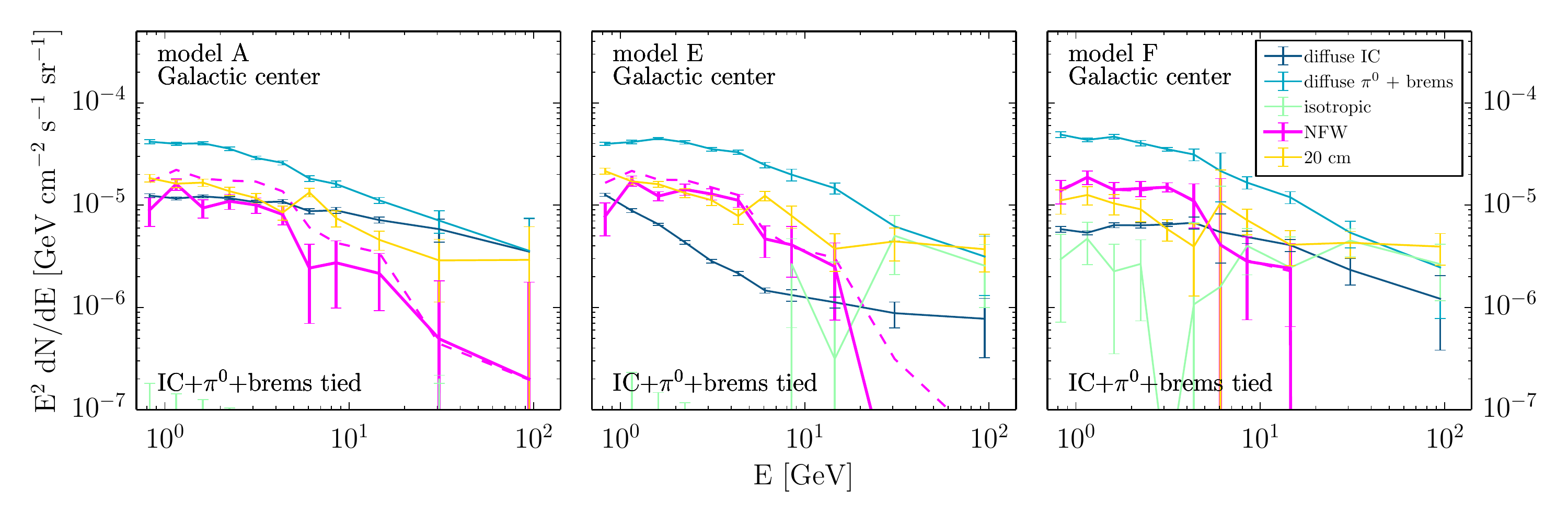}\vspace{-10mm}
 \caption{Same as the top row of Fig.~\ref{fig:bgspectra}, but with the fits performed with the IC and $\pi^0$+bremsstrahlung templates constrained to have the same relative normalizations to each other as predicted by GALPROP. The dashed magenta line plots the GCE spectrum from Fig.~\ref{fig:bgspectra} (where the IC and $\pi^0$+bremsstrahlung were fit separately) for comparison. Note that the dashed comparison NFW annihilation spectrum is indeed plotted in the model F panel but is difficult to see because of its close overlap with the solid magenta NFW spectrum (where the diffuse templates are fixed relative to each other).}
 \label{fig:bgspectra_fulldiffusetied}
\end{figure}

The dropout of the diffuse galactic center IC background below 10 GeV raises the concern that the under-modeling of this component might be causing photons from this source to be falsely attributed to the GCE, and that a significant portion---if not all---of the GCE in the galactic center is simply misattributed IC background emission. 
To test whether this is the case here, we combine the IC and $\pi^0$+bremsstrahlung template into a single template which matches the GALPROP prediction for each component. We then repeat the bin-by-bin template fitting in the galactic center ROI using this single IC+$\pi^0$+bremsstrahlung diffuse template. The results of these fits are shown in Fig.~\ref{fig:bgspectra_fulldiffusetied}. 
By constraining the IC background component to be fixed to its predicted intensity relative to the bright, more easily-fit $\pi^0$+bremsstrahulung component, we are able to recover a physically realistic spectrum for the IC background in the galactic center. 

We see in Fig.\ref{fig:bgspectra_fulldiffusetied} that fitting with a combined IC+$\pi^0$+bremsstrahlung diffuse template causes the GCE spectrum to change at most by a factor of two downwards compared to the case where the IC template normalization was allowed to vary freely. We therefore caution that there may be some degeneracy between the GCE and GALPROP IC components, depending on the chosen background model. The comparison of the GCE spectrum with and without the IC+$\pi^0$+bremsstrahlung templates tied is also shown in Fig.~\ref{fig:bgspectra_fulldiffusetied}, where the dashed magenta line is the NFW annihilation template spectrum from Fig.~\ref{fig:bgspectra}.

In our inner galaxy fits, we find that the GCE spectrum has a more pronounced bump as well as a slightly higher peak normalization at $\sim$2 GeV in fits where the model E diffuse background was used. In the bottom row of Fig.~\ref{fig:bgspectra} we see that this bump feature in the GCE spectra is accompanied by a corresponding dip in the IC diffuse background at the same energies (relative to the best-fit IC spectrum in the other model backgrounds). This suggests that the GALPROP-generated spatial templates for the IC diffuse background at energies $\lesssim$2--3 GeV are a very poor description of the true IC emission in the inner galaxy---so much so that the likelihood fitting procedure finds that a large fraction of the GALPROP-predicted IC emission is better fit by the NFW template than the model E IC template.

\afterpage{
\begin{figure}[p]
 \centering
\includegraphics[width=6.3truein]{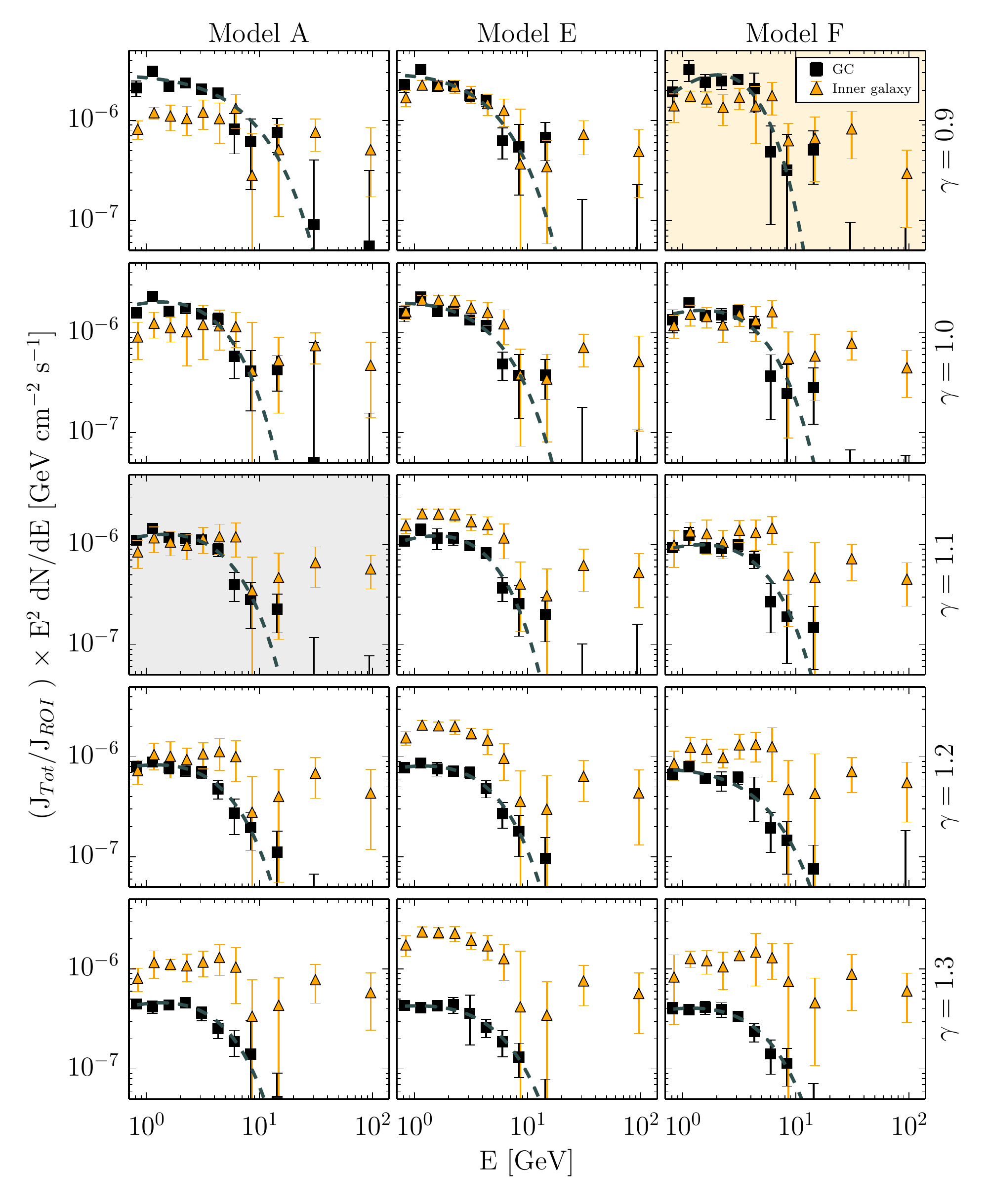} \vspace{-7mm}
\caption{Best-fit GCE spectra in the galactic center (black squares) and inner galaxy (orange triangles) regions, shown for varied GALPROP diffuse models (rows) and NFW density profile slopes $\gamma$ (columns). The spectrum of the GCE in the inner galaxy is shown for the sum of all inner galaxy ROIs. \textbf{Normalizations are scaled to show the expected flux for the entire GCE template ($35^\circ\times35^\circ$)}, such that the normalization for the two ROIs will match if they are consistent with originating from a single NFW-distributed source. Also shown are the exponential cutoff parameterized fits (gray dashed line) to the galactic center spectrum. The panel with the light gray (orange) background denotes the NFW slope and diffuse background combination with the highest likelihood fit for the galactic center (inner galaxy) ROI as recorded in Tab.~\ref{tab:gammaTSvals}. }
\label{fig:GCvsIG_flux}
\end{figure}
\clearpage
}

\subsection{The GCE spectrum in the galactic center versus the inner galaxy}
\label{subsec:GCvsIGspectrum}

Fig.~\ref{fig:GCvsIG_flux} shows the GCE spectrum in the galactic center versus the combined inner galaxy ROIs for all combinations of background diffuse models and NFW templates. We highlight the panels in Fig.~\ref{fig:GCvsIG_flux} which correspond to the highest likelihood background model and NFW template combinations as recorded in Tab.~\ref{tab:gammaTSvals} for the galactic center (gray) and inner galaxy (light orange). The flux in each region is scaled by the J-factor of the entire NFW template divided by the J-factor of the plotted ROI. Thus, all subplots show the expected flux for the entire GCE template ($35^\circ\times35^\circ$), which allows for easier comparison between different regions: if both the galactic center and inner galaxy are consistent with a single NFW-like source, then their data points should have the same normalization in Fig.~\ref{fig:GCvsIG_flux}. With this scaling applied, it is apparent that \textit{for the best-fitting GALPROP backgrounds and NFW profile slopes, the peak intensity of the observed GCE spectrum in the galactic center and inner galaxy regions is consistent with originating from a single NFW source}.

It is also evident from Fig.~\ref{fig:GCvsIG_flux} that the shape of the GCE spectrum in both the galactic center and inner galaxy ROIs is remains consistent throughout the various combinations of GALPROP backgrounds and NFW profiles used in this analysis. The spectrum in the galactic center agrees with the results of previous studies confined to the innermost few degrees of the Milky Way \cite{Abazajian:2014fta, Abazajian:2014hsa, Gordon:2013vta}, where the GCE had a steep cutoff before $\sim$10--20 GeV. \textit{In contrast to our galactic center results, the GCE in the inner galaxy does not exhibit any spectral cutoff and still shows significant flux at energies $\gtrsim$10 GeV. This high-energy tail, as referenced in Sec. \ref{sec:intro}, is also robust to model variations and is present in all combinations of diffuse backgrounds and NFW templates tested here.} We further discuss the significance of this finding in Sec.~\ref{subsec:spectrumtail}.

\subsection{The GCE spatial profile and radial distribution}
\label{subsec:gammas}

Tab.~\ref{tab:gammaTSvals} gives the change in NFW test statistic value (TS=$-2\Delta$ln$\mathcal{L}$) for each combination of GALPROP diffuse model and NFW template slope. The differences in the TS values are given relative to the model with the highest TS value for the NFW template. In the galactic center ROI, we find that the data is best fit with GALPROP diffuse background A and NFW template slope $\gamma$=1.1. However, if using the typical cut of TS$>$25 to determine significance, $\gamma$=1.1 is not significantly favored over $\gamma$=1.2 in the galactic center. Within the inner galaxy, the highest GCE TS values correspond to fits using GALPROP model F. The NFW template in the inner galaxy favors \textit{shallower} profiles with  $\gamma\leq$0.9 when fitting with the model F diffuse background. However, we find that $\gamma$=1.1 is the best fit NFW template in the inner galaxy when using the less-favored background models A and E. \textit{We thus conclude that the slope of the NFW density profile is poorly constrained in the inner galaxy, and variations in diffuse background modeling can have large effects on the best-fit NFW profile slope in that region.}

\begin{figure}[t]
 \centering
\includegraphics[width=6.3truein]{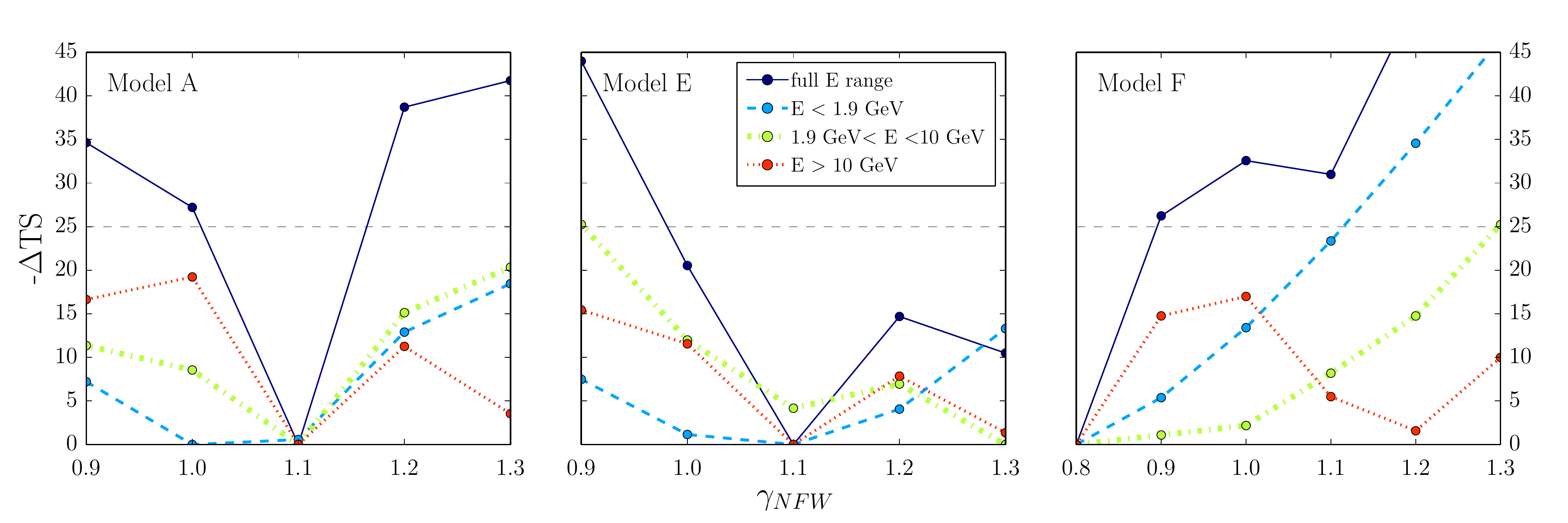}  \vspace{-7mm}
\caption{Change in test statistic value for inner galaxy fits as a function of NFW profile slope $\gamma$. Individual panels show results obtained using GALPROP diffuse models A/E/F. Each line shows $-\Delta$TS calculated using the fits in energy bins $<1.9$ GeV (blue dashed line), 1.9-10.0 GeV (green dot-dashed line), $>10.0$ GeV (red dotted line) as well as the full energy range (solid indigo line). For a given GALPROP background, this allows provides a visualization of which energy bins are driving the fit towards the preferred NFW slope.}
\label{fig:deltaTS_gammas}
\end{figure}

In Fig.~\ref{fig:deltaTS_gammas} we plot the change in TS value as a function of NFW slope $\gamma$ for each GALPROP diffuse model fit in the inner galaxy. The total $-\Delta$TS is broken down into its contributions from energy bins below 1.9 GeV, 1.9--10 GeV, and above 10 GeV. For inner galaxy fits using GALPROP diffuse model F, we see that the preference for shallow profile slopes is most strongly driven by the low energy end of the GCE spectrum below 2 GeV.

\begin{table*}[b]
\centering
\ra{1.3}
\begin{tabular}{@{}lccccccccccc@{}}\toprule& \multicolumn{3}{c}{Galactic center} & \phantom{abc}& \multicolumn{3}{c}{Inner galaxy} \\
\cmidrule{2-4} \cmidrule{6-8}& model A & model E & model F && model A & model E & model F \\ \midrule
$\gamma_{\text{NFW}}$\\
\textbf{0.8} & \dots & \dots & \dots     && \dots & \dots & 0.0 \\
\textbf{0.9} & -44.0 & -81.1 & -112.4     && -349.6 & -976.1 &-21.2 \\
\textbf{1.0} & -14.1 & -49.4 & -105.2      && -342.1 & -952.7 & -32.6 \\
\textbf{1.1} & 0.0 & -33.6 & -101.7     && -314.9  & -932.1 & -31.0 \\
\textbf{1.2} & -4.5 & -36.2 & -103.1     && -353.6 & -946.8 & -50.9 \\
\textbf{1.3} & -34.2 & -63.5 & -119.7     &&  -356.7  & -942.6  & -81.8  \\
\bottomrule
\end{tabular}
\caption{Relative differences in the test statistic (TS) value of the NFW template, given for all combinations of NFW template/diffuse background in both the galactic center and combined inner galaxy ROIs. The differences in test statistic values are given relative to the combination of NFW density slope and diffuse background with the highest TS values: $\gamma$=1.1/model A in the galactic center ROI and $\gamma$=0.8/model F in the inner galaxy ROIs. }
\label{tab:gammaTSvals} 
\end{table*}

\begin{figure}[b]
\centering
\includegraphics[width=6.2truein]{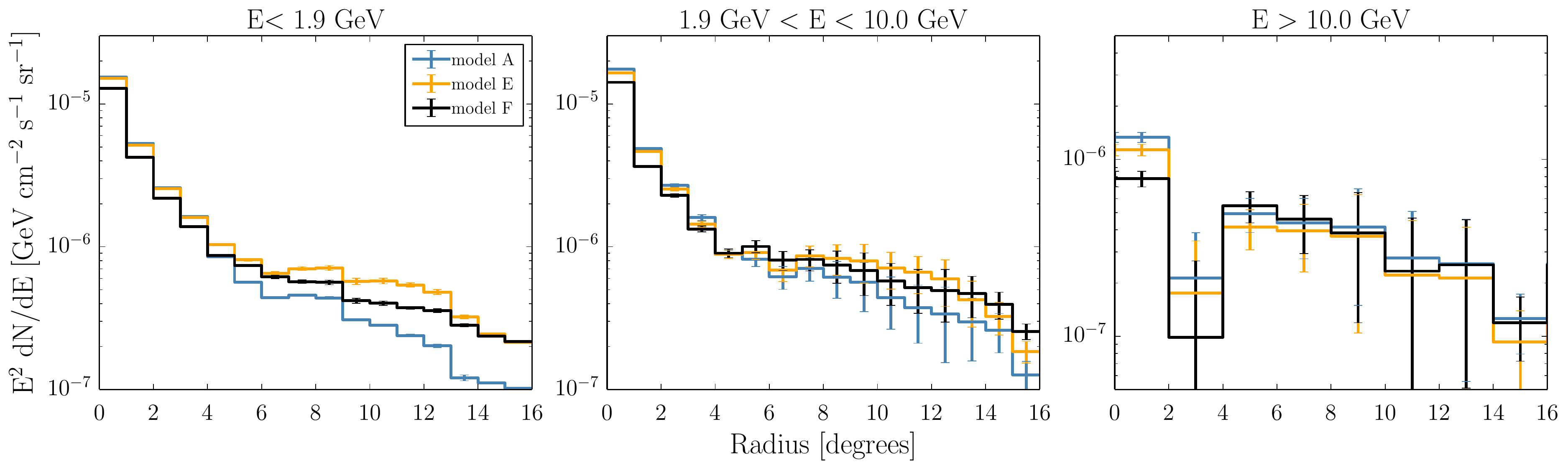}  \\ \vspace{-4mm}
\includegraphics[width=6.2truein]{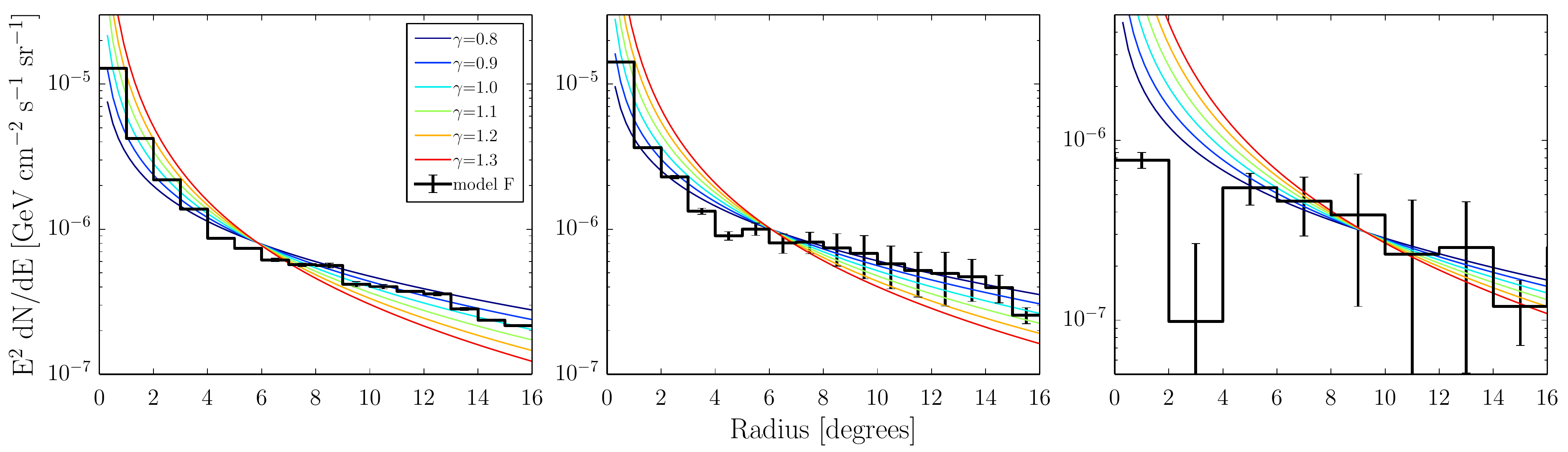} \vspace{-6mm}
\caption{Radial distribution of the GCE residual flux in energy bins 0.7--1.9 GeV, 1.9--10.0 GeV, and 10.0 GeV--200 GeV. The top row plots the GCE residuals for the best fitting profile slopes obtained when fitting with diffuse model backgrounds A/E/F. For all models and ROIs this corresponds to $\gamma=1.1$, with the exception of the inner galaxy model F fit which is plotted for $\gamma=0.9$. The bottom row plots the GCE residual for the fits with $\gamma=1.1$ (0.9) in the galactic center (inner galaxy) and model F diffuse background (solid black steps) against the expected radial distribution of emission from NFW-like sources of varying profile slopes (colored lines) comparison. }
\label{fig:radial_distrib}
\end{figure}

Our weak constraint on the NFW profile slope $\gamma$ in the inner galaxy is seemingly in contrast with the findings of previous works which strongly favor spatial profiles for the GCE with $\gamma\sim$1.1--1.3 in the inner galaxy, with little to no dependence on background modeling \cite{Daylan:2014rsa,Calore:2014xka}. We attribute this discrepancy to differences in the minimum galactocentric distance used by various authors to define their region of interest. Ref.~\cite{Calore:2014xka} derive their constraints on $\gamma$ by analyzing the region defined by $2^\circ \leq b \leq 20^\circ$ and $l \leq 20^\circ$, while Ref.~\cite{Daylan:2014rsa} use $1^\circ \leq b \leq 20^\circ$ and $l \leq 20^\circ$. Our `galactic center' ROI---within which we find that $\gamma$ is consistently 1.1--1.2 for all GALPROP diffuse models---overlaps with the regions in these works between latitudes of $1^\circ-2^\circ \leq b \leq 3.5^\circ$. As we show in the top row of Fig.~\ref{fig:radial_distrib}, the radial profile of the excess below 10 GeV is largely insensitive to changes in the GALPROP background model out to $r\sim5-6^\circ$.

Fig.~\ref{fig:radial_distrib} shows the radial distribution of flux in the GCE-associated residual for energies below 1.9 GeV, 1.9--10.0 GeV, and above 10 GeV. The top row of Fig.~\ref{fig:radial_distrib} plots this quantity for all three GALPROP backgrounds with the best-fit NFW profile residuals in the galactic center and inner galaxy. Below 10 GeV, the radial profile of the GCE within a $6^\circ$ radius shows little variation with changes in background modeling. The systematic effects associated with the diffuse modeling become more apparent outside of this radius, especially below 1.9 GeV, where the differences between the radial profiles for fits with backgrounds A/E/F are larger than the error bars of each radial bin.

In the bottom row of Fig.~\ref{fig:radial_distrib} we plot the expected radial profiles for NFW haloes with density slopes of $\gamma=0.8-1.3$ along with the observed radial profile of the best-fit GCE residuals for GALPROP model F. The normalizations of the NFW profiles are adjusted to best fit the entire radial range of GCE residual data points for energy bins below 1.9 GeV and from 1.9--10.0 GeV. For the energy bin above 10.0 GeV, the curves for varying NFW profiles are fit to the inner galaxy data points ($r\geq5^\circ$). No single NFW template is able to fit all the radial bins; for example, between 1.9--10 GeV, all the NFW profiles tend to over-predict flux between $4^\circ - 5^\circ$, with shallower (steeper) profiles under(over)-predicting flux at lower radii and over(under)-predicting flux at higher radii. At energies above 10 GeV, the radial profile of the GCE residual is decidedly non-NFW-like due to a drop in flux within $4^\circ$. This is consistent with our earlier observation that the GCE spectrum in the galactic center drops off steeply around 10 GeV, while its spectrum has a high-energy tail in the inner galaxy ROIs.

\subsection{Spatial uniformity of the galactic center excess spectrum}
\label{subsec:spatialGCE}

In Fig.~\ref{fig:GCvsIG_flux} we identify a high-energy tail in the GCE spectrum which is present in combined inner galaxy ROIs (as seen in Refs.~\cite{Calore:2014xka, TheFermi-LAT:2015kwa}), but not in the galactic center ROI. Here, we compare the spectra between individual inner galaxy ROIs (as defined in Tab.~\ref{tab:ROIs}) to explore the spatial uniformity of the GCE spectrum and its high-energy tail.

\begin{figure}[t]
 \centering
\includegraphics[width=6.4truein]{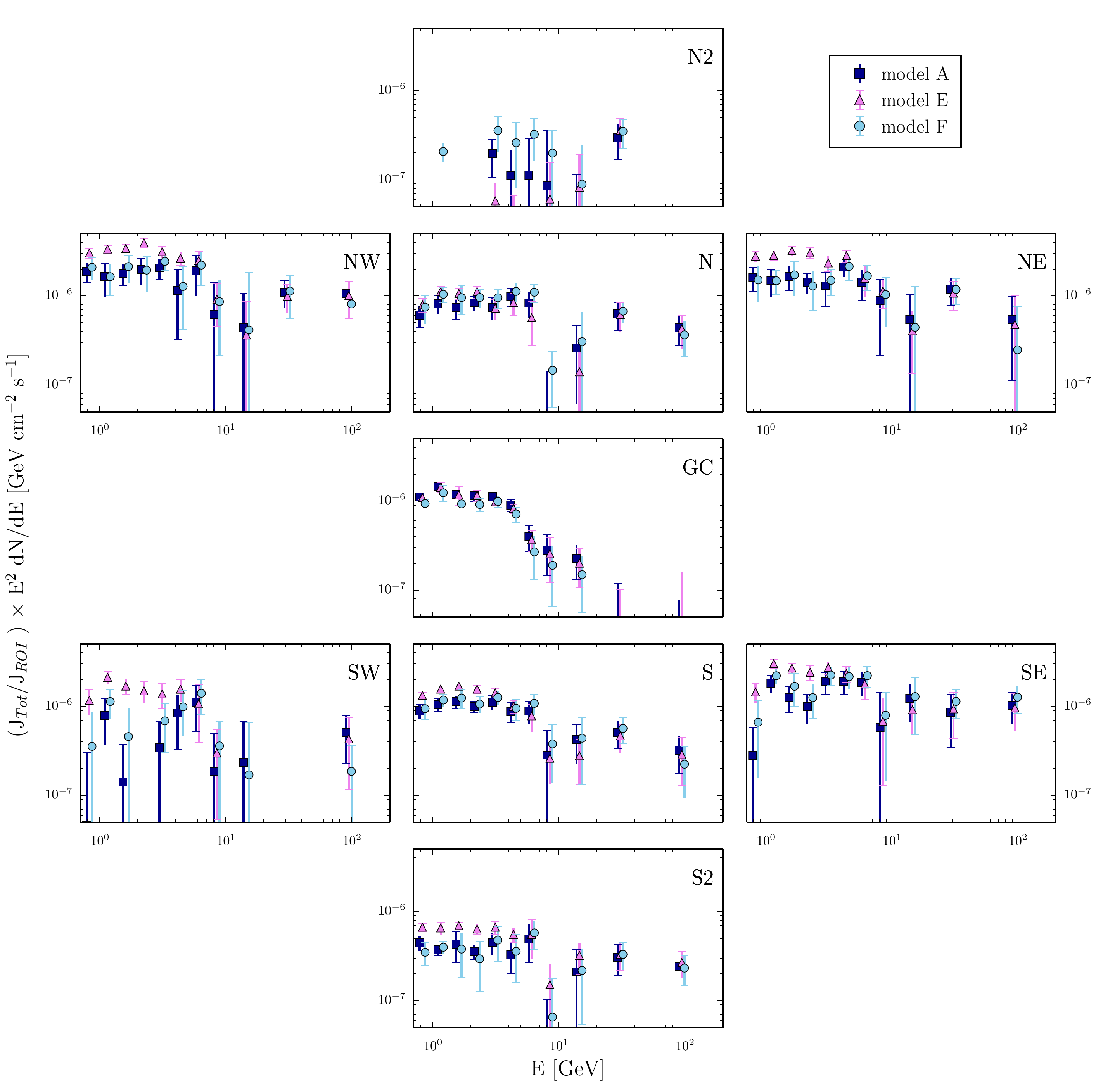}

\caption{Best-fit GCE spectra in the galactic center ROI and the 6 inner galaxy ROIs, shown for fits with diffuse backgrounds model A (dark blue), E (purple), and F (light blue). Additional regions N2/S2 are also plotted. Panels are arranged to reflect location on the sky.  We show our results for the case of a GCE template with profile slope $\gamma=1.1$, though our findings our consistent for all template slopes. Normalizations are scaled such that each subplot shows the expected flux for the entire GCE template ($35^\circ\times35^\circ$). Energy coordinates of the data points are slightly offset for visibility purposes.}
\label{fig:IG_ROIs_3x3}
\end{figure}

Fig.~\ref{fig:IG_ROIs_3x3} shows the spectrum of the GCE-associated residual (observed counts$-$ full model + best fit GCE model) for the subregions defined in Fig.~\ref{fig:ROIs}. Also included are two additional ROIs N2 and S2 (defined in Tab.~\ref{tab:ROIs}) which lie at farther latitudes ($11^\circ < |b| < 19^\circ$) than the N/S ROIs used in the inner galaxy analysis. We plot these additional ROIs in Fig.~\ref{fig:IG_ROIs_3x3} to check (1) the approximate spatial extent of the GCE and (2) whether the GCE spectrum in the farther-latitude regions, if detected, is consistent with the spectrum at closer galactocentric radii. Note that the N2 and S2 regions are not included in the analysis of the GCE in the inner galaxy. The error bars are shown for the statistical uncertainties in the binned fluxes, while the systematic uncertainties associated with the background modeling may be roughly estimated by the spread in the GCE spectrum fit with the different background models.

The spectrum in each ROI subplot is normalized in the same way as in Fig.~\ref{fig:GCvsIG_flux}, where the GCE flux is scaled by the J-factor of the entire NFW template divided by the J-factor of the smaller ROI. The average normalizations between $\sim$1--5 GeV in the GC and the separate inner galaxy ROIs (with the exception of the SW ROI) are in agreement within a factor of two. If taken at face value, this difference in GCE normalizations across the separate ROIs may be interpreted as a rough estimate of the uncertainty in the axis ratio of the GCE's projected spatial distribution. Interestingly, the normalizations in the NW/NE/SE ROIs are higher on average than those in the N/S ROIs (although still overlapping within statistical and systematic uncertainties), which may hint at some degree of compression along the longitudinal axis in the GCE spatial profile.

We are unable to identify any specific source or extended feature in the SW ROI as the cause of this discrepancy. It is a strong possibility that the differences in the SW GCE spectrum are due to mismodeling of the diffuse background in that region, as the IC templates calculated by GALPROP are symmetric in $l$ and $b$ and are thus unable to capture the non-axisymmetric variations in the true background diffuse emission. 

At farther distances from the galactic center, the GCE spectrum in the S2 ROI appears roughly similar in shape to the GCE spectrum in the inner galaxy, but has a slightly lower overall normalization. The GCE spectrum in the N2 ROI is consistent with zero flux for two out of the three diffuse backgrounds tested. These results may indicate that the GCE's intensity profile is well-described by an NFW profile out to angular distances of $\sim 10-12^\circ$, outside of which it falls off more steeply. However, the expected flux from an NFW profile at radii $>10^\circ$ is quite low and thus any determination of the GCE spectrum at these larger radii is subject to great uncertainty.

Within the framework of this analysis, we find that \textit{the high-energy tail of the GCE is a large-scale spatial feature that is present in all of the inner galaxy ROIs, with the exception of the SW ROI.} We previously noted that the high-energy spectrum in the inner galaxy is in sharp contrast to the galactic center, where no GCE emission is observed at energies $\gtrsim$10--20 GeV. In Fig.~\ref{fig:IG_ROIs_3x3} we show that (with the exception of the SW ROI) the GCE spectrum in all inner galaxy ROIs falls by a factor of roughly 2--3 between its peak at $\sim2$ GeV and the highest energy bin (44.7--200.0 GeV). The high-energy GCE emission is prevalent throughout the inner galaxy, and is not a result of one region heavily biasing the combined inner galaxy spectrum. 

It is possible that the high-energy tail of the GCE spectrum is simply misattributed flux from the Fermi bubbles. This hypothesis is supported by the absence of the tail in the low-latitude galactic center ROI and the similarity of the hard bubble spectrum to the GCE spectrum above 10 GeV (both are roughly $\sim$E$^{-2}$). The bubble morphology becomes uncertain at low latitudes, and may perhaps cover a larger fraction of the inner galaxy ROIs than assumed in our template \cite{Fermi-LAT:2014sfa,MalyshevFermiTalk}. We consider it unlikely that the Fermi bubbles are responsible for the majority of the high-energy GCE flux if we assume that the rough bubble template used in this analysis is a reasonable approximation for the extent of the bubbles in the inner galaxy ROIs. As can be seen in Fig.~\ref{fig:ROIs}, the fraction of each individual ROI in the inner galaxy covered by the bubble template ranges from $\sim$0.1 (SE) up to 1.0 (N). Even if the bubble template used was not fully accurate in tracing the bounds of the bubbles, we would still expect to observe varying normalizations in the high-energy tail between different ROIs if these photons were in fact originating from the Fermi bubbles (i.e. the spectrum above 10 GeV would have the highest intensity in the N ROI and the lowest in the SE ROI). We do not observe any correlation between the intensity of the spectrum above 10 GeV in Fig.~\ref{fig:IG_ROIs_3x3} and the fraction of each ROI overlapping with the bubble template.

\section{Discussion}
\subsection{The GCE high-energy spectrum above 10 GeV}
\label{subsec:spectrumtail}

In Figs.~\ref{fig:GCvsIG_flux} and \ref{fig:IG_ROIs_3x3} we explore the spatial dependence of the GCE spectrum and confirm previous findings by Refs.~\cite{Calore:2014xka, TheFermi-LAT:2015kwa} of GCE-associated emission upwards of $\sim20$ GeV in the inner galaxy region. We find that this high-energy tail is not present in the GCE spectrum within the galactic center ROI; within approximately $5^\circ$ it has a spectral cutoff between $\sim$5--10 GeV, while outside of this region  the high-energy tail becomes a prominent feature for $r\gtrsim7^\circ-8^\circ$. This spectral feature (or lack thereof) is robust to variations in the density profile of the NFW template and persists through variations in the diffuse $\pi^0$ decay + bremsstrahlung and IC background templates. 

This difference between the high-energy GCE spectra in the galactic center versus the inner galaxy may be construed as either (1) a systematic effect associated with uncertainties in the background modeling, (2) an intrinsic spatial variation in the source contributing to the GCE spectrum, or (3) the signature of multiple sources with different spatial profiles. We do not find evidence of the former in this work, as the high-energy inner galaxy GCE spectrum above 10 GeV is recovered in all the GALPROP models and GCE spatial templates we used (Fig.~\ref{fig:GCvsIG_flux}). Ref.~\cite{TheFermi-LAT:2015kwa} do find that the GCE spectrum in the innermost $15^\circ\times15^\circ$ shows a dependence on background modeling: their `index-scaled' diffuse emission models result in a softer GCE spectrum above 10 GeV than their `intensity-scaled' models. When fit to an exponential cut-off functional form, the GCE spectrum cuts off before 10 GeV in the index-scaled background fits; however, when the GCE spectrum is fit as a power law in individual energy bins, the power law-like high-energy tail is present for all background models (albeit with a softer index in for the index-scaled cases).

Here we will assume that (1) is not the case and discuss what implications might follow for interpretations of the galactic center excess as either dark matter annihilation or an astrophysical source. Under this assumption, the $\gtrsim20$ GeV emission implies that a MSP population or dark matter source would need to produce both a prompt gamma-ray component (peaking in the $\sim$GeV range) as well as a hard leptonic component (which produces the high-energy tail above 10 GeV through IC scattering).

\subsubsection{Dark matter annihilation and the GCE spectrum}
We will first consider the case of dark matter annihilation producing the GCE. The simplest dark matter annihilation models fit the excess with prompt gamma-ray emission\footnote{Here, `prompt' emission refers to the gamma rays produced through the hadronization and/or decays of the primary annihilation products as well as higher order corrections to the dark matter annihilation diagrams.}, without the need for the primary annihilation products to produce secondary gamma-ray emission through the environment-dependent processes of IC or bremsstrahlung scattering. Prior to claims of the excess emission extending beyond $\sim$10 GeV in energy, the GCE spectrum was most commonly fit with WIMPs annihilating into $\sim$10 GeV $\tau$ leptons or $\sim$40 GeV b quarks \cite{Goodenough:2009gk, Hooper:2010mq, Hooper:2011ti, Abazajian:2012pn, Gordon:2013vta, Macias:2013vya, Abazajian:2014fta, Calore:2014xka}. However, the gamma-ray spectra of these oft-mentioned $\tau^+ \tau^-$ and $b\overline{b}$ annihilation modes cut off sharply by about 10--20 GeV, which is difficult to reconcile with the inner galaxy GCE spectrum we observe beyond those energies (although consistent with the GCE in the galactic center). Producing a high-energy tail through prompt dark matter annihilation alone is still possible, but requires models such as prompt annihilation of WIMPs into nonrelativistic Higgs (m$_\chi \simeq$ 126 GeV) \cite{Calore:2014nla, Agrawal:2014oha}. 

The prompt gamma-ray spectrum is only dependent on the particle physics involved in the dark matter annihilations and subsequent Standard Model hadronizations and/or decays. Assuming these processes are independent of environment, we would expect the GCE spectrum to be spatially uniform if it was due to prompt dark matter annihilation. In contrast, we observe that the GCE spectrum has a power law-like tail at high-energies in the inner galaxy but not the galactic center. If this discrepancy is a true feature of the GCE and not a systematic error, it disfavors the interpretation of the GCE as emission from prompt dark matter annihilation.

Alternatively, the GCE could be produced through secondary emission from Standard Model annhilation products, as discussed in Sec.~\ref{sec:intro}. Secondary emission from IC or bremsstrahlung processes is dependent on the environment and may result in a spatially varying GCE spectrum. Previous authors have fit the GCE spectrum with IC scattering off of dark matter annihilation products \cite{Kaplinghat:2015gha,Calore:2014nla}. However, neither of these proposed IC scenarios are capable of producing a high-energy tail at larger galactocentric radii while suppressing it at lower radii because the IC target density decreases with distance from the galactic center. 

In order for dark matter annihilations to reproduce the spatial variation we observe in the high-energy GCE spectrum, the spectrum of the secondary e$^+$e$^-$'s would need to have a cutoff energy \textit{higher} than that of the primary annihilation products producing the prompt emission. The simplest dark matter annihilation models that fit the GCE with a combination of prompt and IC emission from a single annihilation channel (e.g. prompt annihilation into muons accompanied by IC scattering of the secondary e$^+$e$^-$ for m$_{\chi}\simeq$ 60--70 GeV \cite{Calore:2014nla}) have secondary e$^+$e$^-$ spectra with cutoff energies below that of the primaries. Dark matter annihilation through multiple channels including direct annihilation into electrons may produce a harder electron spectrum, but the branching ratios and cross sections for that channel are tightly constrained by AMS-02 limits on electron-positron spectral lines features \cite{Bergstrom:2013jra,Bringmann:2014lpa}.

Our arguments outlined above consider only single-component dark matter models. It may be possible to explain both the excess and its spatial variation through a model with two dark matter particles, with the higher-mass particle annihilating preferentially into leptons. We also note another possibility that the high-energy tail arises from an astrophysical source while the bulk of the excess below 10 GeV is due to dark matter annihilation. In this case the dark matter interpretation is not expected to be significantly different from that considered in the early papers \cite{Goodenough:2009gk, Hooper:2010mq, Hooper:2011ti, Abazajian:2012pn, Gordon:2013vta, Macias:2013vya, Abazajian:2014fta} and so we do not discuss it further here.

\subsubsection{Leptonic cosmic-ray outbursts and the GCE spectrum}
Given that the Fermi bubbles are evidence of an extremely energetic outburst in the Milky Way's past, it is possible that the GCE may originate from one or more cosmic-ray outbursts \cite{Carlson:2014cwa,Petrovic:2014uda,Gaggero:2015nsa, Cholis:2015dea}. Interpretations of the GCE as the product of burst events tend to focus on models dominated by leptonic, rather than hadronic, cosmic rays, as gamma-ray emission following a hadronic outburst would be strongly correlated with the gas distribution in the plane of the disk and thus would not be consistent with the approximately spherical morphology of the GCE \cite{Carlson:2014cwa}. 

The cosmic-ray electron spectrum changes with distance from the galactic center due to diffusion and energy losses. The combination of a spatially varying electron spectrum and interstellar radiation field should lead to a similarly non-uniform GCE spectrum. In this regard, our finding of a radially varying GCE spectrum would seem to support the leptonic outburst scenario. However, it is still difficult to explain why the GCE high-energy tail would be present in the inner galaxy but not the galactic center, as cosmic-ray propagation outwards from a central source would result in the opposite effect. If the GCE is due to multiple outbursts, the harder spectrum at farther radii might be produced through a burst with a considerably harder injection spectrum than the more recent bursts which contribute to the excess at lower radii.

An additional complication in modeling the GCE with cosmic-ray outbursts is replicating the steeply rising GCE spatial profiles within small radii. The angular profiles of cosmic-ray GCE models are flat at low radii, and multiple recent fine-tuned bursts (within a few hundreds of years) are needed to produce a GCE profile that continues to rise at low radii \cite{Cholis:2015dea}.

\subsubsection{Millisecond pulsars and the GCE spectrum}

The similarity of the $\sim$2 GeV bump in the GCE spectra reported in Refs.~\cite{Hooper:2011ti, Abazajian:2012pn,Gordon:2013vta,Macias:2013vya,Abazajian:2014fta,Abazajian:2014hsa, Daylan:2014rsa} to Fermi observations of resolved MSPs and globular clusters (which host populations of millisecond pulsars) motivates the interpretation of the signal as emission from an unresolved population of MSPs concentrated at the galactic center \cite{Abazajian:2010zy,Abazajian:2012pn,Mirabal:2013rba,Petrovic:2014xra,Yuan:2014rca,Yuan:2014yda,Lacroix:2015wfx,O'Leary:2016osi}. Ref.~\cite{O'Leary:2016osi} also discuss the possible contribution to the GCE from young pulsars. We will focus this discussion on MSPs as the potential unresolved young pulsar population is concentrated in the plane of the disk and thus would not produce the roughly spherical profile of the GCE.

\begin{figure}[h]
 \centering
\includegraphics[width=3.truein]{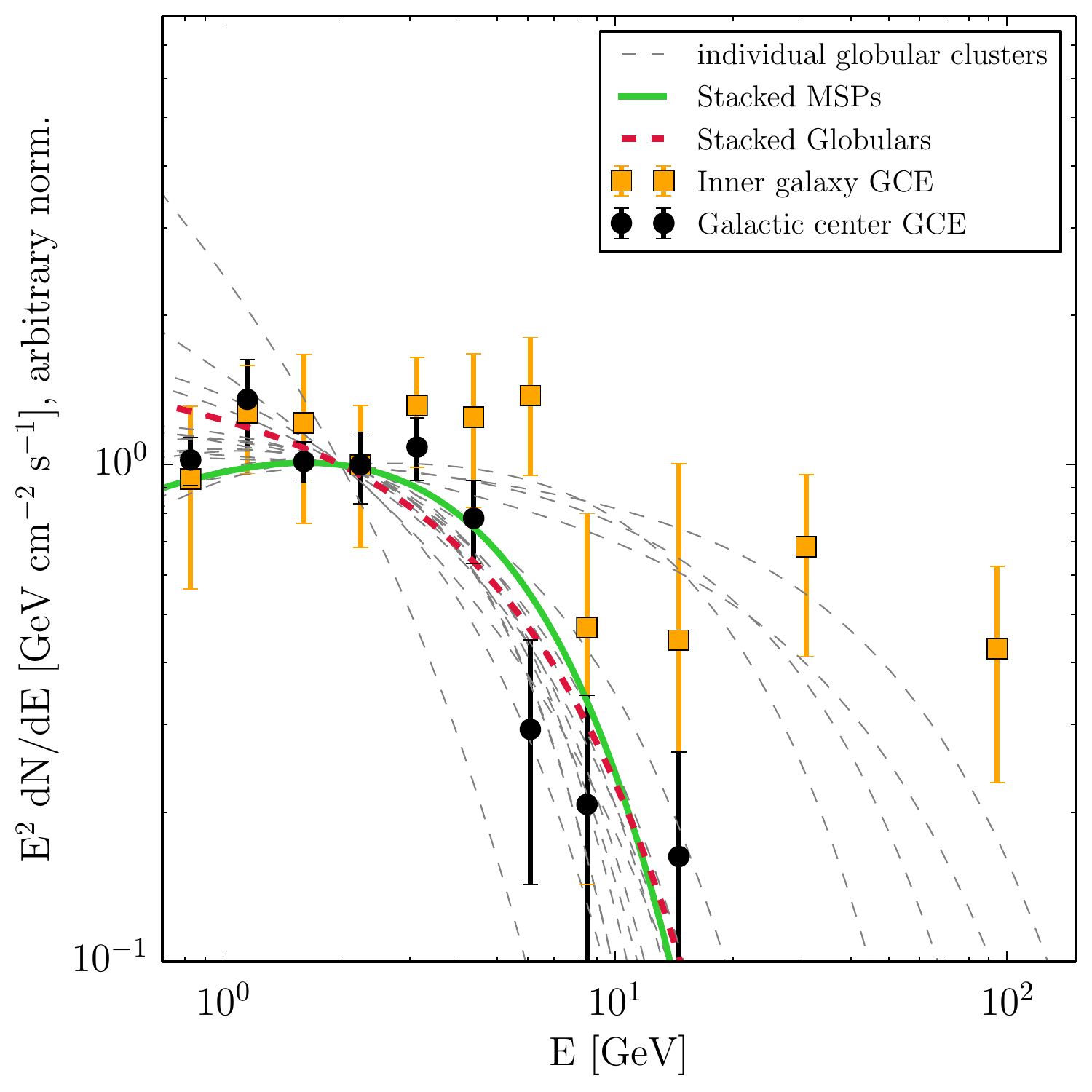}
\includegraphics[width=3.truein]{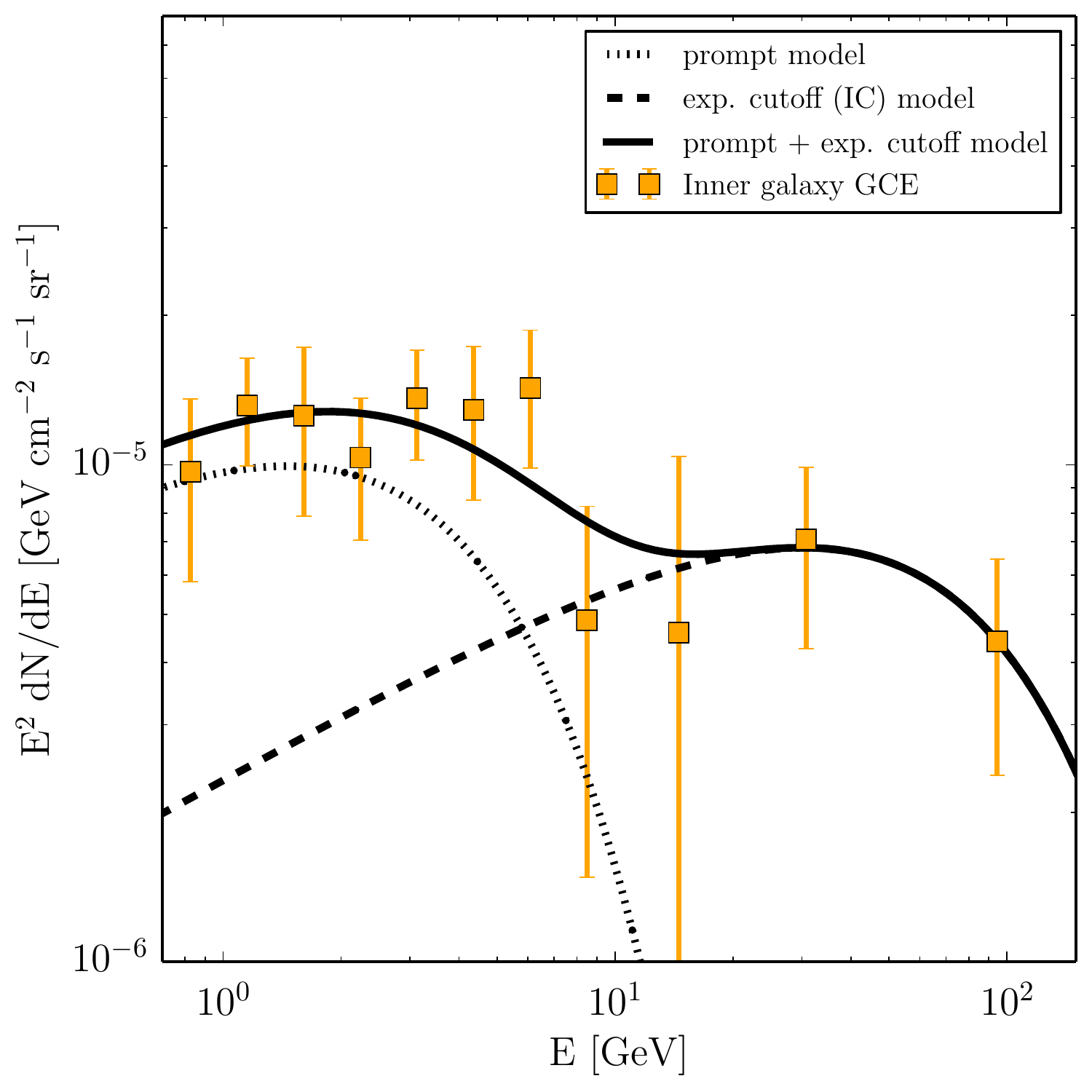}
\vspace{-6mm}
\caption{Left: Comparison of the GCE spectrum in the galactic center (black circles) and inner galaxy (orange squares) to the spread of globular cluster MSP spectra (dashed gray lines) from Ref. \cite{Cholis:2014noa}. Also shown are the inferred MSP spectra from Ref.~\cite{Cholis:2014noa} derived from stacked observations of globular clusters (solid green line) and individual MSPs (dashed red line). The spectra are normalized to match each other at 2 GeV for ease of comparison. Right: Comparison of the inner galaxy GCE spectrum (orange squares) to our estimations of the prompt (dotted line) and IC (dashed line) components of the potential unresolved MSP population.}
\label{fig:MSPcompare}
\end{figure}

In the left panel of Fig.~\ref{fig:MSPcompare} we compare the GCE spectrum in both the galactic center and inner galaxy ROIs to the prompt spectra of MSPs as measured in Ref.~\cite{Cholis:2014noa}. Although the spectral shape of the excess in the galactic center is consistent with the spectra of the stacked globular clusters and stacked individual MSPs, the high-energy tail of the GCE spectrum in the inner galaxy is a distinct departure from the sharp cutoffs at $\lesssim$10 GeV in the stacked spectra. The inner galaxy GCE spectrum is barely consistent with prompt MSP emission---while the typical globular cluster or MSP spectrum cuts off before 10 GeV, there are a handful of globular clusters (M5, M62, NGC 6624, NGC 6752) whose parameterized spectra predict gamma-ray emission at energies above 15 GeV. Thus, one could claim that the high-energy tail of the GCE in the inner galaxy is not entirely inconsistent with prompt MSP emission, if using the outliers in the globular cluster sample as a comparison. However, these clusters also have very large uncertainties in their fitted spectra: the $68\%$ lower confidence interval on the spectral cutoff energy lies below 15 GeV for all of the aforementioned outliers. 

Furthermore, \textit{it may be possible for a MSP population to produce the variation in the high-energy tail through a combination of prompt emission from the MSPs themselves as well as secondary IC emission from the e$^+$e$^-$ injected by the MSPs into the interstellar medium.} Ref. \cite{Petrovic:2014xra} point out that for certain cosmic-ray propagation parameters in their models, the secondary IC emission from MSPs is subdominant to the prompt signal within $\sim2^\circ$ in latitude but becomes comparable to---or even greater than---the prompt emission at latitudes outside this range. If the electron injection spectrum has a high cutoff energy above $\sim100$ GeV, the secondary IC emission would extend beyond 10 GeV and could give rise to the prominent high-energy feature in the inner galaxy. A comparison of the GCE spectra in the galactic center and inner galaxy ROIs may in fact be suggestive of this: the inner galaxy GCE spectrum resembles a composite of the galactic center spectrum with a harder IC $\sim E^{-2}$ power law spectrum extending past 100 GeV.

We explore this possibility in the right panel of Fig.~\ref{fig:MSPcompare} by plotting the inner galaxy GCE spectrum against the combined prompt and IC components that might arise from an unresolved MSP population. The `prompt' component of this model is taken to be the best-fit exponential cutoff parameterization of the GCE in the galactic center (for GALPROP model A and $\gamma=1.1$) with a freely floating normalization. The `IC' component is fit as an exponential cutoff spectrum with the normalization, index, and cutoff energy free to vary. The combination of the best-fit `prompt+IC' spectrum is shown as the heavy solid line in the right-hand side of Fig.~\ref{fig:MSPcompare}. 

The best-fit IC parameterization has a power-law index $\Gamma=1.55$ and a cutoff energy of 68 GeV, indicating that a MSP origin would require a very hard spectrum spectrum of outgoing electrons with energies up to $\mathcal{O}(10)$ times greater than the maximum energy of the prompt gamma-ray emission. Such an injection spectrum may be achievable through one of several mechanisms proposed to accelerate MSP electrons to energies $>$100 GeV. Ref.~\cite{Harding:2011mw} find that offsets between the polar cap and magnetic dipole axis in MSPs can produce e$^+$e$^-$ cascade pairs with energies up to $\mathcal{O}(100)$ times greater than in young pulsars; however, the outgoing pair spectra in these models have softer power-law indices than the gamma-ray IC fit in Fig.~\ref{fig:MSPcompare} might require. A harder $\sim$TeV electron spectrum may be produced through reacceleration of e$^+$e$^-$ at intrabinary shock fronts within MSP systems \cite{Venter:2015gga} or in pulsar wind shocks \cite{Bednarek:2007nn,Bednarek:2013oha,Yuan:2014yda}.

The large uncertainties in MSP electron injection spectra leave room for a hard e$^+$e$^-$ population to produce both the GCE spectrum above 10 GeV and its spatial variation. An unresolved population of MSPs is therefore a compelling explanation for the GCE with (1) the prompt gamma-ray emission dominating the spectrum between $\sim1-5$ GeV, accounting for the similarity between the galactic center and inner galaxy GCE spectra in this energy range and (2) the spatially dependent IC emission dominating the inner galaxy spectrum above $\sim20$ GeV. Future work (in progress) will model the combined prompt+IC spectrum from a $\sim r^{-(1.6-2.2)}$ MSP distribution as a function of galactocentric distance and investigate how this population contributes to the WMAP haze.\footnote{Ref.~\cite{Kaplinghat:2009ix} study \textit{young} pulsars as a possible source for the WMAP haze; however, it is hard to reproduce the latitudinal extent of the haze with the disk-like young pulsar distribution.}

Radio and microwave observations provide some constraints on leptonic emission from MSPs, but are highly dependent upon the parameters assumed for the cosmic-ray injection and propagation \cite{Cholis:2014fja,Bringmann:2014lpa,Egorov:2015eta}. Radio observations utilizing the upcoming Square Kilometer Array (SKA) will provide greater power to detect more MSPs in addition to any synchotron emission associated with their secondary IC emission \cite{Smits:2008cf,Calore:2015bsx}. If the GCE high-energy tail is indeed IC radiation from an electron population with $\gtrsim$TeV energies, the possible extension of this component into photon energies above the Fermi-LAT sensitivity range may be detectable by next-generation of TeV-scale gamma-ray observatories such as the Cherenkov Telescope Array (CTA) \cite{Yuan:2014yda}.

Of course, the GCE might originate from a combination of multiple astrophysical sources. The inner $\lesssim5^\circ$ of the GCE may be due to prompt emission from an unresolved population of MSPs (which produce the spectrum with a $\sim10$ GeV cutoff that we observe in the galactic center) while the dominant contribution to the harder gamma-ray GCE spectrum at larger radii comes from one or more leptonic cosmic-ray bursts. In this combined scenario, the GCE spatial profile in the galactic center arises because of the $\sim r^{-2.2}$ distribution of the putative MSP population, and there is no need to invoke a series of recent outbursts to explain the steep rise in the centralmost regions.

\subsection{Is a point source population favored over a smooth annihilation profile as the origin of the GCE?}
\label{subsec:ptsrcs}

Recent works have attempted to determine whether the GCE originates from a smoothly distributed NFW annihilation source or a population of $\mathcal{O}$(1000) faint point sources with fluxes below the Fermi-LAT detection sensitivity. Ref.~\cite{Bartels:2015aea} employ a wavelet decomposition analysis of Fermi-LAT data and find that the photon clustering structure is compatible with the estimated radial distribution and spectrum of a faint MSP population. Ref.~\cite{Lee:2015fea} use non-Poissonian photon-count statistics to differentiate between the signal produced by a smooth NFW source versus a unresolved point source distribution. Their analysis favors a point source origin for the GCE, where all of the excess might be explained by a source-count distribution with a sharp decline just below the Fermi detector sensitivity ($\sim1-2 \times 10^{-10}$ ph cm$^{-2}$ s$^{-1}$). 

\begin{figure}[b]
 \centering
\includegraphics[width=6.truein]{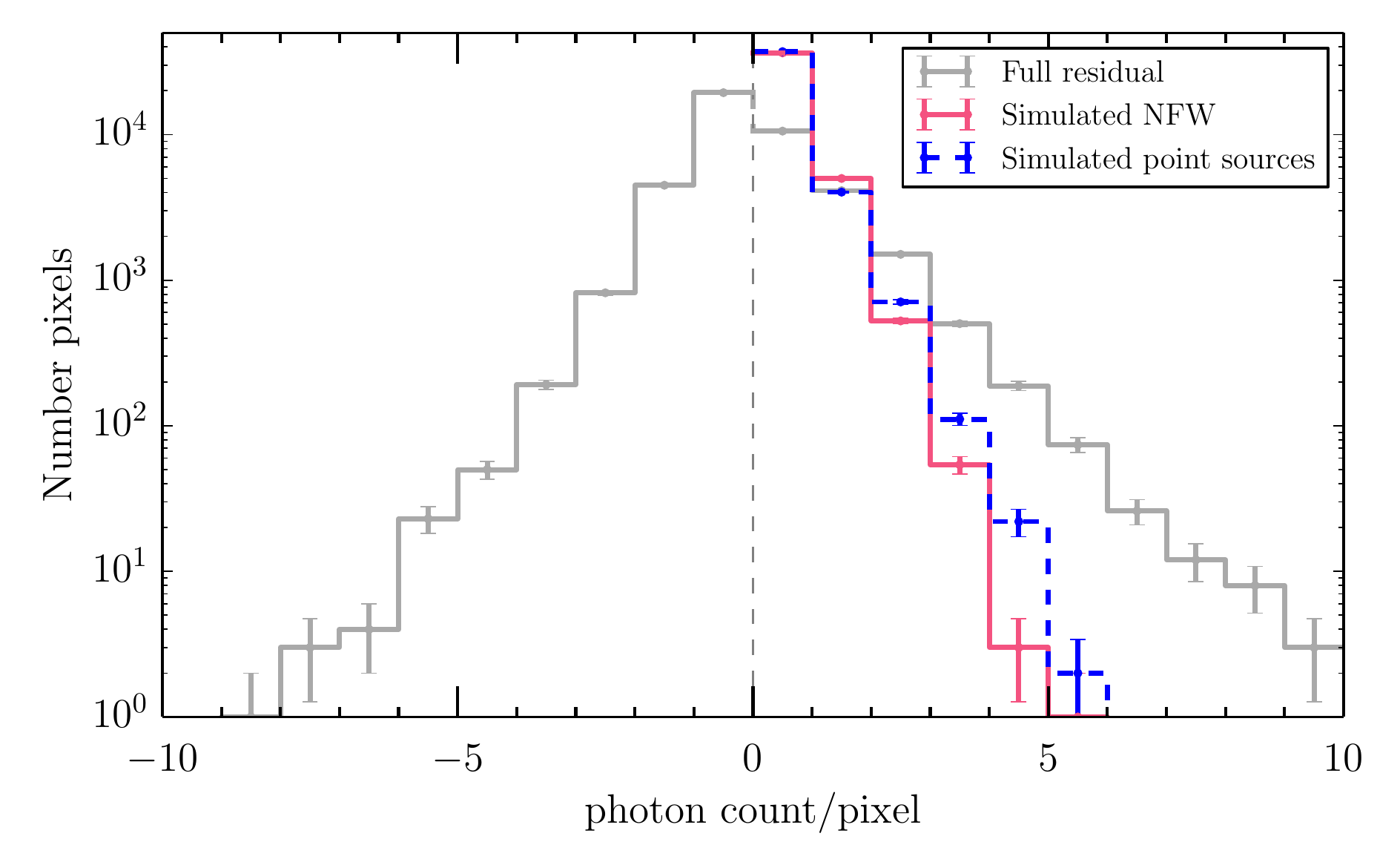}

\caption{Simulated photon-count distribution in the combined inner galaxy ROIs between 1.9--10.0 GeV of an NFW source with $\gamma=1.1$ (dashed blue line) vs. a population of unresolved point sources (solid pink line) with the best-fit source count function from Ref. \cite{Lee:2015fea}. Pixel size is 0.2$^\circ$ per pixel. The spectrum and total flux of each simulated source is chosen to match our total observed flux in this energy range. For comparison, we also show the photon-count distribution of the full residual (solid gray line) for the case of GALPROP model F and $\gamma=1.1$. Note that the zero-count bin is not shown.  }
\label{fig:photoncounts}
\end{figure}

Here, we explore whether the photon-count distribution of the inner galaxy GCE residual allows us to distinguish between the unresolved point sources and dark matter annihilation scenario for the GCE source. We simulate a population of unresolved point sources in our inner galaxy ROIs with a radial distribution $\sim r^{-2.2}$. Point source fluxes are drawn from a source-count function dN/d$\Phi$ [sources/(ph cm$^{-2}$ s$^{-1}$)] modeled by a broken power law of the form $\Phi^{-\alpha_1 (\alpha_2)}$ below (above) the break flux $\Phi_b$, where $\Phi$ is the flux [ph cm$^{-2}$s$^{-1}$] per point source between 1.9--11.9 GeV. We use the parameters \{$\Phi_b=2.16\times10^{-10}$ ph cm$^{-2}$ s$^{-1}, \alpha_1=-0.57, \alpha_2=29.5 $\} from the source-count function found by Ref.~\cite{Lee:2015fea} to be the best fit point source population. The spectrum and total flux of the simulated point source distribution is chosen to match the inner galaxy GCE spectrum for $\gamma=1.1$ and the model F diffuse background. We find that $\sim$1900 point sources are required within an 18$^\circ$ radius to match our inner galaxy GCE spectrum and flux using this source-count function. About 300 of these sources lie within a 10$^\circ$ radius with $|b|>2^\circ$, which corresponds to the region within which Ref.~\cite{Lee:2015fea} require $203^{+109}_{-68}$ point sources to explain the excess.

We use the \textit{gtobssim} tool~\cite{fermitools} to simulate photon events originating from (1) the point source population described above and (2) annihilation signal from an NFW source with $\gamma=1.1$ and the same spectrum as the point sources. The observations are simulated using the same timeframe and cuts as described in Sec.~\ref{sec:methods} and include the effect of the instrument's energy-dependent PSF. In Fig.~\ref{fig:photoncounts} we plot the distribution of photon-counts per pixel for pixels with non-zero counts. We show both of the simulated observations, as well as the photon-counts per pixel for the full residuals of the fit using $\gamma=1.1$ and the model F background.  

As expected, the simulated observations (solid lines) in Fig.~\ref{fig:photoncounts} show that the smooth annihilation source has more pixels with low photon-counts (1 ph/pixel) while the point source distribution has more pixels with higher photon-counts ($\geq$2 ph/pixel). The statistical method described in Ref.~\cite{Lee:2014mza} takes advantage of this difference to determine whether a point source population or smooth NFW halo is the true source of the GCE. 

By comparison, we find that the amplitudes of the positive and negative pixel-count distributions for the full residuals (gray lines) are an order of magnitude larger than the differences between the point source population and NFW halo annihilation. Over- and under-subtractions in the gamma-ray residuals may be due to multiple issues in the modeling, and are not necessarily due to real gamma-ray emitting features. We do not presume to understand the underlying causes of the large positive and negative fluctuations in our residuals.\footnote{For a thorough description of the intricacies of modeling the diffuse emission towards the galactic center, see Ref.~\cite{TheFermi-LAT:2015kwa}. We note that the residual counts per pixel area shown in Ref.~\cite{TheFermi-LAT:2015kwa} are of similar order to this work, despite their use of specialized background modeling tuned to flatten the residuals.} Lacking an understanding of the effects that give rise to the photon-count distribution of the residuals, it is possible that small-scale spatial structure in the mismodeling may be erroneously interpreted as sub-threshold point sources. Our analysis of the photon-count distribution is not sufficiently sophisticated to estimate the extent to which this mismodeling may affect current sub-threshold point source analyses, but it provides a visual demonstration of this systematic uncertainty.

\section{Conclusions}

We find that the inclusion of an extended, spherically symmetric gamma-ray source with an NFW-like radial profile of $\sim r^{-(1.6-2.2)}$ strongly increases the fit likelihood obtained through the template fitting procedure within $\sim10^\circ-15^\circ$ of the galactic center. The galactic center excess spectrum obtained through the likelihood template fitting procedure is reasonably robust to variations in the NFW and background diffuse model templates used in this work, even in extreme cases where the background modeling is likely an unphysical description of the true extended gamma-ray sources. These findings are in agreement with many previous studies of the excess~\cite{Goodenough:2009gk,Vitale:2009hr, Hooper:2010mq,Hooper:2011ti, Abazajian:2012pn,Gordon:2013vta,Macias:2013vya,Hooper:2013rwa, Abazajian:2014fta,Abazajian:2014hsa,Zhou:2014lva, Daylan:2014rsa,Calore:2014xka,TheFermi-LAT:2015kwa,Cumberbatch:2010ii}.
If we compare the galactic center and inner galaxy spectra for $\gamma=1.1$, the peak normalizations are consistent in both ROIs.
 
When varying the GALPROP-generated IC and $\pi^0$+bremsstrahlung background models, we find that the galactic center ROI within $\lesssim4^\circ-5^\circ$ is consistently best fit with an NFW profile slope around $\gamma=1.1$. Outside of this radius, however, the best fit NFW profile in the inner galaxy is poorly constrained ($\gamma\lesssim0.8-1.1$), with a heavy dependence on the choice of diffuse background model (Tab.~\ref{tab:gammaTSvals}). Our results suggest that previous works \cite{Daylan:2014rsa,Calore:2014xka}, in which the profile slope was found to be well-constrained to $\gamma\sim1.1-1.2$ when using a $\sim20^\circ$ ROI about the galactic center,
may have been driven strongly by the inclusion of the innermost few degrees of the excess during the template fitting procedure.

The most noticeable difference between the GCE in the galactic center versus the inner galaxy regions is the hardening of its spectrum at galactocentric radii above $\sim5^\circ - 6^\circ$ (Figs.~\ref{fig:GCvsIG_flux}, \ref{fig:radial_distrib}). In the inner galaxy, we observe a power law-like tail in the GCE spectrum extending upwards of 100 GeV, while the spectrum in the galactic center has a steep falloff at $\sim$10 GeV. 

The inner galaxy high-energy tail above 10 GeV is found in all but one of the inner galaxy ROIs and is robust to variations in the diffuse background models and the GCE spatial templates used in this work. This presence and intensity of this high-energy component is roughly consistent across most of the inner galaxy ROI and shows no obvious azimuthal asymmetry (Fig.~\ref{fig:IG_ROIs_3x3}).
Upon examination of the radial distribution of GCE photons above $\sim$10 GeV, we see that this high-energy spectral feature is roughly consistent with an NFW annihilation profile outside of a $\sim5^\circ-6^\circ$ radius, but does not exhibit the steep rise in brightness towards lower radii that we observe for the GCE photons below 10 GeV. 

If the full energy range of the GCE emission in the centralmost few degrees as well as the outlying regions is assumed to arise from a single source, then a single component dark matter annihilation model cannot account for the spatial variation of the high-energy GCE emission above 10 GeV.
Of course, it is possible that the high-energy tail and bulk of the excess below 10 GeV are due to two different sources, in which case there is no difficulty in explaining the excess below 10 GeV as arising from dark matter annihilation. 

We attempt to use the photon-count distribution of the GCE residual to distinguish between the scenarios of dark matter annihilation in a smooth NFW halo and an unresolved population of MSPs. However, this effort is inconclusive as the amplitude of the full residuals is greater than the GCE amplitude by a factor of $\sim$few in almost all photon count bins (Fig.~\ref{fig:photoncounts}). We thus caution that the residual photon count distribution resulting from mismodeling of the data may be a confounding factor when using photon count statistics to search for point source populations.

Although we are unable to confirm the existence of an unresolved MSP point source population, it remains a compelling explanation because of its close match with the GCE spectrum below 10 GeV and the potential for the MSP population to produce a secondary leptonic component at energies significantly higher than that of the prompt gamma-ray emission. If the MSP electron injection spectrum is sufficiently hard and extends upwards of $\sim100$ GeV, the GCE emission above 10 GeV may be attributable to IC scattering of these high-energy electrons. The  spatial variation of the high-energy tail of the GCE spectrum described in this work would then follow as a natural result of electron propagation. 
Looking towards the future, our understanding of the true source(s) of the GCE will be greatly advanced by combining multiwavelength observations with the ongoing efforts involving realistic modeling of cosmic-ray propagation\footnote{As this manuscript was being prepared we became aware of Ref.~\cite{Carlson:2016iis}, which studies the dependence of the GCE spectra upon a set of physically-motivated background diffuse models. There are some similarities in results between Ref.~\cite{Carlson:2016iis}, particularly with regards to the potential degeneracies between the GCE and GALPROP IC components as well as the background-dependence of the best-fit GCE morphology.} and novel statistical analyses.

\acknowledgments
We thank Kevork Abazajian, Sheldon Campbell, Oscar Macias and Simona Murgia for useful discussions and comments, as well as Farhad Yusef-Zadeh for sharing the 20 cm radio map used as a template in this work. M.K. is supported by NSF Grant No. PHY-1316792. A.K. is supported by NSF GRFP Grant No. DGE-1321846.

\bibliography{master}

\end{document}